\documentclass[useAMS,usenatbib]{mn2e}
\usepackage{latexsym,hhline,epsfig,longtable,amsmath,amssymb}
\usepackage{graphicx,color}  
\usepackage{ulem}

\def\apj{ApJ}
\def\apjs{ApJS} 
\def\aap{A\&A}
\def\apss{Ap\&SS}

\def\mnras{MNRAS}

\def\apjl{ApJL}

\def\araa{ARA\&A}

\def\aj{AJ}

\defcitealias{EDL06}{EDL06} 

\def\chem#1{$^{#1}$} 
\def\msun{M$_\odot$} 
\def\lsun{L$_\odot$}
\def\rsun{R$_\odot$}
 
\def\Lsun{L$_\odot$}

{}

\newcommand{\pdc}[3]{ \left( {{\partial #1}\over{\partial #2}}\right)_{#3} }

\title[On the Numerical Treatment and Dependence of Thermohaline Mixing in Red Giants]{On the Numerical Treatment and Dependence of Thermohaline Mixing in Red Giants}

\author[J. C. Lattanzio et al.]{J. C. Lattanzio$^{1}$\thanks{E-mail:john.lattanzio@monash.edu}, L. Siess$^{2}$, 
       R. P. Church$^{3,1}$, G. Angelou$^{1}$, R. J. Stancliffe$^{4}$, 
\newauthor C. L. Doherty$^1$, T. Stephen$^{1}$ and  S. W. Campbell$^{1}$\\
$^{1}$Monash Centre for Astrophysics (MoCA), School of 
Mathematical Sciences, Monash University, Victoria 3800, Australia\\
$^{2}$Institut d'Astronomie et d'Astrophysique, Universit\'e Libre de Bruxelles (ULB), CP 226, B-1050 Brussels, Belgium\\
$^{3}$Lund Observatory, Department of Astronomy and Theoretical Physics, Box 43, SE-221 00 Lund, Sweden\\
$^{4}$Argelander Institute for Astronomy, University of Bonn, Auf dem Huegel 71, D-53121 Bonn, Germany}

\begin{document}
\normalem
\maketitle

\begin{abstract}
In recent years much interest has been shown in the process of thermohaline mixing in red giants.
In low and intermediate mass stars
this mechanism first activates at the position of the
bump in the luminosity function, and has been identified as a likely
candidate for driving the slow mixing inferred to occur in these stars.
One particularly important consequence of this process, which is
driven by a molecular weight inversion, is the destruction of lithium.
We show that the degree of lithium destruction, or in some cases production, 
is extremely sensitive to the numerical details of the
stellar models. Within the standard 1D diffusion approximation to thermohaline mixing, we find that
different evolution codes, with their default numerical schemes, can produce lithium abundances that
differ from one another 
by many orders of magnitude. This disagreement is worse for faster mixing. We perform experiments with four
independent stellar evolution codes, and derive conditions for the spatial and temporal resolution
required for a converged numerical solution. The results are extremely sensitive
to the timesteps used. We find that predicted lithium abundances published in the literature until now
should be treated with caution.
\end{abstract}


\section{Introduction}

Lithium remains an enigmatic element. Because it captures a proton at such low 
temperatures (about 2 million K) we find that stars are much more efficient 
at destroying Li than producing it. It is 
synthesized by cosmic ray spallation and was made only in trace amounts in the Big Bang.  
In 1982 it was discovered that there is a baseline minimum Li abundance 
in old stars in the galactic halo. This is  known as the ``Spite plateau'' 
after the discoverers \citep{SpitePlateau}, and corresponds to a value of 
Li/H $\simeq 1.1 \times 10^{-10}$ by number. With this background we are able to
elucidate a number of so-called ``lithium problems''.
\begin{enumerate}
\item There remains a significant discrepancy between the predictions for the Li abundance 
from Big Bang nucleosynthesis and the WMAP results (Li/H $=5.2\times 10^{-10}$ or 
A(Li)=2.716\footnote{A(Li) = $\log \left( n({\rm Li})/n({\rm H})\right) + 12$ where $n$ denotes the number density.}) compared to 
observations in the oldest Population II stars (essentially the Spite Plateau) which 
show a consistent Li/H $\simeq$ 1--2$ \times 10^{-10}$ or A(Li)$=2.00 -2.30$ \citep{Cyburt2008}.
\item There is also a dip in the Spite Plateau appearing over a narrow range of effective
temperatures $T_{\rm eff} \simeq 6400$--$6900$ (corresponding to dwarf masses in the range $\simeq 1.2$--$1.5$\msun). 
This requires some form of mixing to transport the Li to regions where it is destroyed, but which only
acts over this narrow range in spectral type or mass. For reviews see \citet{Bal1995,Pin97,AntTwa09}.
\item Our Sun has a surface Li abundance about 140 times lower
than predicted by the best models for the early evolution of the Sun \citep[e.g.][]{Meletal2010}.
\item In some (galactic) globular clusters, both populations of stars (usually interpreted as different generations) 
 seem to show the same Li content despite the fact that 
one population shows the results of hot hydrogen burning, which should efficiently destroy any Li present.
This observation thus requires some Li production but, 
curiously, it must occur at exactly the rate required to match the material 
that has not undergone burning \citep{DM2010}. 
\end{enumerate}
Such long-standing and confounding problems present headaches for stellar 
physicists, and make it ironic that Li is used to treat depression in humans.

\begin{figure*}
\resizebox{0.65\hsize}{!}{\includegraphics{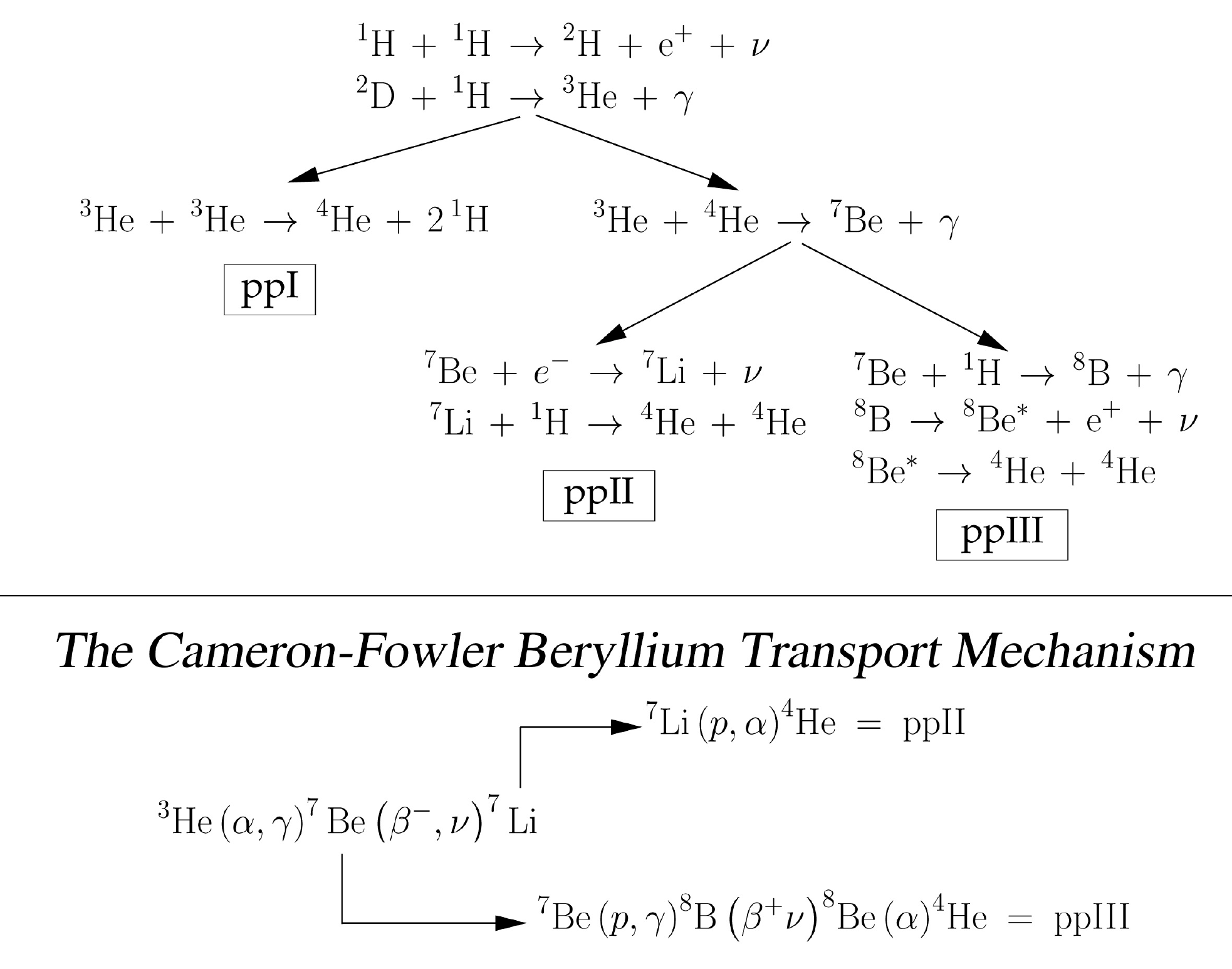}}
\caption{Upper panel: The pp chains. Lower panel: the Cameron-Fowler Beryllium Transport Mechanism.}
\label{reactions}
\end{figure*}

Lithium is involved in the classical pp chains of hydrogen burning, as shown
in Figure~\ref{reactions}. The ppII and ppIII chains start with
the production of \chem{7}Be.  At typical hydrogen burning temperatures the \chem{7}Be 
can capture a proton which leads to the completion of the ppIII chain. 
But one can also produce 
\chem{7}Li through electron capture on \chem{7}Be, which occurs at all temperatures.
This Li is then 
destroyed by a proton capture, completing the ppII chain. However, it was pointed out some time ago by 
\cite{CFBTM} that if one can move the \chem{7}Be to a cooler region before it captures 
a proton, the still rapid electron capture could lead to a net production of \chem{7}Li. 
This is the basis of the ``Cameron-Fowler Beryllium Transport Mechanism''
which is understood to be active in the more massive AGB stars where we find hot-bottom burning. In these
stars the convective envelope extends into the top of the hydrogen burning shell, with the result
that nuclear burning happens in a very thin zone at the bottom of a dynamically active convective zone. Models of this
phase by \cite{SB92Li} and  \cite{Mazz99} seem to explain the observed super-Li-rich AGB stars seen in the Magellanic Clouds 
\citep{SL89Li,SL90Li}.

In recent years it has been realised that during normal evolution along the red-giant branch, 
low mass
stars will develop an inversion in their molecular weight distribution when the hydrogen shell reaches 
the maximum interior extent of the convective envelope during the first dredge-up  episode 
(hereafter FDU) \citep{EDL06}. This mixing is driven
by the inversion in mean molecular weight created when the last reaction in the ppI 
chain takes place. Although a fusion reaction,  it replaces two \chem{3}He nuclei 
with three nuclei - one \chem{4}He and two protons. Thus the mean 
molecular weight $\mu$ decreases. Normally in a hydrogen burning region this 
decrease is more than compensated for by the increase from the other fusion reactions.
But when the hydrogen burning shell has passed the abundance discontinuity left 
by FDU, the far-from-equilibrium \chem{3}He burns at a lower temperature than 
the hydrogen and reduces $\mu$ locally.

This is a normal part of the life of a red-giant and should happen in almost all stars. Studies by
various authors \citep{CZ07, EDL08, CantLang2010, M3paper, M92paper} have shown that the resultant mixing, usually 
referred to as thermohaline mixing in analogy to the case of salt and thermal diffusion in the oceans, seems
to produce the required decrease in the $^{12}$C/$^{13}$C  ratio beyond that predicted by FDU.
It also produces a decrease in \chem{7}Li \citep{CZ07,CL10} as required by observations,
and by destroying $^3$He it produces agreement between Big Bang nucleosynthesis and
observations of the $^3$He content of the Galaxy \citep{EDL06,Charbonnel95,Lagarde12}.
\citet{Stan-etal2009} have shown that it is also able to reproduce the observed abundances
in carbon-normal and carbon-enhanced metal-poor (CEMP) giants.

In the region of this local minimum in $\mu$ we have both \chem{3}He and
\chem{4}He at temperatures at which nuclear reactions are operating, 
and hence we expect \chem{7}Be and \chem{7}Li to be present. 
A quantitative study is needed to determine the overall impact of
this process on the evolution of the envelope composition. Thermohaline
mixing is a prime candidate for
production of \chem{7}Li as well as destruction, and only careful study can
reveal which dominates.
Indeed, we note that there is
a need for some mechanism to produce Li in low mass red giants as many Li-rich giants are found at
luminosities well below that required for hot-bottom burning, which is the usually favoured 
mechanism for Li production \citep{AL96,Petal2011}. \citet{Stan2010} has shown that thermohaline mixing
can match the Li observations in carbon-enhanced metal-poor stars.

In this paper we investigate the sensitivity of Li production and destruction to the numerical details
of the calculation of the thermohaline mixing found on the first giant branch. We show that unless great
care is taken, the same implementation of the mixing phenomenon in different codes can produce results 
that differ by orders of magnitude. We concentrate on Li because of its extreme sensitivity to 
physical conditions in the model. Although this work is done within the framework of the
linear theory for thermohaline mixing as developed by \citet{Ulrich72} and \citet{Ketal80}, the conclusions
will apply to any mechanism that determines the abundance of the fragile element Li. This
is especially true if that process uses the composition to determine the 
efficiency of mixing, as we discuss below. This feedback ensures that care must be taken to
obtain an accurate solution.

\section{Thermohaline Mixing on the Red-Giant Branch}

\citet{EDL06} showed that during the ascent of the first giant branch, red giants would
naturally develop a molecular weight inversion when the hydrogen burning shell reached the abundance discontinuity
left behind by 
the maximum inward extent of the convective envelope during FDU. 
Although this particular application of thermohaline mixing was only recently realised, the 
process itself has been known for a long time. The linear theory has been developed by 
\cite{Ulrich72} and extended to the non-perfect gas case by \cite{Ketal80}. They
have cast the theory into a form suitable for a diffusion equation, although we would
do well to remember that most forms of mixing are advective, not diffusive. Nevertheless, this is a common approximation
in stellar interior studies and it is the one we use here. The thermohaline mixing diffusion coefficient is 
given by
\begin{equation}
D_{thm} = C_t K \left( {{\phi}\over{\delta}} \right) {{\nabla_\mu}\over{\nabla-\nabla_{ad}}}
\end{equation}
where $K$ is the thermal diffusivity given by 
\begin{equation}
K = {{4 a c T^3}\over{3 \kappa \rho^2 c_P}},
\end{equation}
\begin{equation}
\phi = \pdc{\ln \rho}{\ln \mu}{P,T},
\end{equation}
\begin{equation}
\delta = -\pdc{\ln \rho}{\ln T}{P,\mu},
\end{equation}
and all other symbols have their usual meaning. Note that $\phi = \delta = 1$ for a perfect gas,
and that $D_{thm} = 0$ unless $\nabla_\mu < 0$.

The \citet{Ulrich72} and \citet{Ketal80}  formulations are
equivalent, and depend on one unknown parameter $C_t$, a non-dimensional coefficient
theoretically related to the aspect ratio $a$ (length/width) of the mixing fingers.
Again, we caution that this is a highly idealized formulation, but within this model we have \citep{CZ07}
\begin{equation}
C_t = {{8}\over{3}} \pi^2 a^2.
\end{equation}
Comparison with observations has led to the preferred value of $C_t \simeq 1000$ to match
the \chem{12}C/\chem{13}C ratio
seen both in field stars and in globular clusters
\citep{CZ07, EDL08, CL10, M3paper, M92paper}, and also abundances in CEMP stars \citep{Stan-etal2009, Stan2010}.
It is remarkable that a single value produces so much agreement with observations. 
Yet from a theoretical viewpoint the situation is far from satisfactory.

Various researchers have tried to improve our understanding of the mixing process by 
producing numerical simulations of one kind or another. Calculations have been performed with the 
Boussinesq approximation in 2D \citep{Den2010} as well as 3D \citep{DenMer2011,TGS11}; see
also \citet{Brown13}. These
authors all agree that the resulting aspect ratio is too small to support
the value $C_t\simeq 1000$ required to match the observations. Nevertheless, that the observations fit
this theory so nicely is a compelling fact, also pointed out by \citet{Den2010}. 
We note that the mechanism naturally starts at
the correct place on the giant branch, 
and there is no freedom in choosing this as it is determined by the 
extent of first dredge-up and the temperature dependence of the
pp chain reactions. 
So what are we to make of the seeming contradiction between the hydrodynamical simulations
and the 1D models? There are three things to consider: 
\begin{enumerate}
\item how reliable are the calculations that use the linear theory?
\item how reliable are the determinations of $C_t$ from observations?
\item how reliable are the hydrodynamical models?
\end{enumerate}
The first question is the topic of this paper, but let us briefly address the other two questions.

It is true that a value of $C_t \simeq 1\,000$ seems to be required to match many observational
constraints, as listed earlier. However recent work by \citet{Angelou14} has shown that a 
value of $C_t \simeq 100$ is required to match observations of Li in various globular clusters. 
This is an order of magnitude smaller, and is getting closer to the values preferred by
the existing hydrodynamical simulations. So which is correct: is it 100 or 1000? Surely 
we cannot choose a different value for 
each species. At this stage we cannot say anything more definite. The thermohaline
mechanism is under investigation from many different approaches and much is uncertain.
For example, the linear theory assumes that a diffusion equation describes the mixing. But if it
is more advective then this may not be appropriate, and the temperature history of a parcel of
stellar material may be important. This may reveal itself as different values of $C_t$ 
for species with different sensitivities to the temperature.
Much work remains to be done; the details are simply unknown at present.

As for the hydrodynamical simulations themselves, these are very difficult calculations.
Most are not performed under stellar conditions \citep{TGS11,Brown13}, and use extrapolations or 
asymptotic behaviour to try to extend the applicability. This is in contrast to the work
of \citet{Den2010} and \citet{DenMer2011} who did simulate stellar conditions. 
\citet{Brown13} found that their simulations did not support the linear
theory, but agreed with those of \citet{Den2010} although they believe the 
latter are under-resolved.
We also note that \citet{Den2010}
estimated that a value of $C_t \simeq 800\,000$ would be required to find Li production. We show
below that this can be achieved for $C_t = 10\,000$, indicating again the need for 
extreme care in performing these calcualtions.

Finally, we refer to a recent paper by \citet{Medrano14} who investigated the hydrodynamics
of thermohaline mixing in the presence of a horizontal inhomogeneity. They found that
the efficiency of the mechanism was increased by orders of
magnitude over the previous hydrodynamical estimates, and noted that this may bring the
hydrodynamical calculations closer to successes of the linear theory.

Clearly the situation is far from settled.
In the rest of this paper we will use the linear theory of \cite{Ketal80},
assuming that $C_t$ takes values consistent with the observations,
and we will look at the resulting predictions from different codes for the same
theory. Debate about the correct theory belongs elsewhere - for now we are concerned 
only with the accurate numerical solution of the existing 1D theory in different stellar 
evolution codes. We repeat that although $C_t$ is formally related to
the aspect ratio of the expected fingering convection (see equation 5), 
this association relies on
the highly simplified idealized case. We prefer to treat $C_t$ as an arbitrary 
scaling parameter in the theory, much like $\alpha_{MLT}$ in the mixing-length theory.

It is prudent to note that any competing theory to thermohaline mixing that is intended to
produce the required abundance changes in red giant stars, must operate over the
same physical region in the star and hence its implementation would require the same
sort of numerical care that we discuss below. Such stringency is required by the
fragility of lithium, rather than any specific mixing mechanism. This is exacerbated if
the proposed mixing mechanism is dependent on the composition profile, as is the case in
thermohaline mixing. It is the feedback from the composition on the mixing that makes the
solution so sensitive to the numerical considerations we discuss below. This is likely
to be true for other mechanisms also.

\section{The Codes Used and the Case Tested}
We will compare calculations made with four different evolution codes, and 
two different versions of the same code, before and after a rewriting of the 
difference scheme  and solution method. All calculations are made initially 
with each code's standard criteria 
for space and time resolution, so that no special demands are enforced. The results 
thus reflect what one might expect for a naive calculation from each code. Salient features
of the codes are described below. We note that the nuclear reactions involved are quite standard
pp chains, and that there have been no significant changes in their rates for years, so we
do not expect any differences to arise thereby. Convective borders are determined
using the normal Schwarzschild criterion, with no overshoot. Any variations from this 
are discussed in the subsections below, dealing with each code.

\subsection{\sc MONSTAR}
The first code we will use is {\sc monstar}, the stellar structure code developed at 
Monash University. This code began life as the Mt Stromlo code \citep{WZ81} but has been 
modified extensively since. We use the OPAL opacity from \citet{OPAL96} without 
allowing for changes in the envelope composition, which are small in the cases considered here.
We use a mixing-length parameter $\alpha_{MLT} = 1.75$ from fitting the Sun to the standard mixing-length theory.
The solutions for the structure and the chemical composition (as a result of burning and 
mixing) are separate in this code. However, it is perhaps best to think of them as interleaved, 
because the latest structure iteration is used for determining the burning rates and mixing regions 
that are applied at the current iteration. At each iteration during the convergence process 
we update the composition by calculating the burning and mixing.  
In this way both structure and composition should 
converge to a self-consistent solution at the same time.

We will see below that the timestep is crucial to the calculations we are investigating.
We note 
that {\sc monstar} has multiple criteria that must be met. Broadly speaking these limit
the timestep so that changes in dependent physical and composition variables (at each mass shell), and 
the total luminosity (from various sources) are below specified limits. 

\subsection{\sc STARS}

The {\sc stars} code was originally written by \citet{PPE71,PPE72},
and has been subsequently updated by many authors \citep[e.g.][]{Polsetal1995}. 
The version used here is that described in \citet{StaEld2009}, 
which is publically available from the {\sc stars} website - see \texttt{http://www.ast.cam.ac.uk/$\sim$stars/}. 
It also uses the OPAL opacities from \citet{OPAL96}, with a mixing length of 2.00 based on calibration to the Sun. 
The solution of the compositions of species involved in energetically significant nuclear reactions
are solved simultaneously with 
the structure equations \citep{RJS-simult}. The implementation of thermohaline mixing in the code is 
described in \citet{Stan-etal2009} and \citet{Stan2010}. 
We note that the calculation of the mean molecular weight $\mu$ is, in this code, 
made with the composition from the {\it previous\/} timestep. This is significant for
the problem addressed in this paper, as we will see below.
Alongside the main evolution routines, the code 
has a set of nucleosynthesis subroutines that follow the evolution of 40 nuclear species 
\citep{Setal-2005}. It is these routines that are used to calculate the lithium abundance.
The timestep is controlled by restricting the sum of the changes in all variables at all mesh points
to be equal to a specified value; there is also a restriction on the maximum change allowed in any 
dependent variable at any timestep.

\subsection{\sc MESA}
We also make use of the recently developed {\sc mesa} suite of stellar evolution codes - {\tt http://mesa.sourceforge.net} and 
described in \cite{MESA1,MESA2}; specifically we use {\sc mesa} version 6208. 
The formulation for thermohaline mixing in the {\sc mesa} code  is slightly different with \citep{MESA2}:
\begin{equation}
D_{thm} = \alpha_{th} {{3k}\over{2\rho c_P}} {{B}\over{\nabla-\nabla_{ad}}}
\end{equation}
where $k$ is the thermal conductivity given by 
\begin{equation}
k = {{4 a c T^3}\over{3 \kappa \rho}},
\end{equation}
and $B$ is given by equation~6 in \citet{MESA2}. Referring to \citet{CantLang2010} we see that
\begin{equation}
B = \left( {{\phi}\over{\delta}}\right) \nabla_\mu
\end{equation}
and hence 
\begin{equation}
C_t = {{3}\over{2}} \alpha_{th}.
\end{equation}
We use the Ledoux criterion for convection, 
and $\alpha_{MLT} = 1.6$ in the \citet{BV} implementation of the  mixing-length theory.
{\sc mesa} 
selects its timestep via a two stage process \citep{MESA1}. 
A timestep is calculated using a scheme based on digital control theory,
which is then subjected to a wide range of tests 
that reduce this timestep if certain properties are changing faster than specified.
Such tests include limits similar to those discussed above for {\sc monstar} and {\sc stars}, 
including restrictions on changes in the solution mesh, composition, nuclear burning rates, etc.

\subsection{\sc STAREVOL-GENEVA-V2.3}
Our final code is actually two variants of the {\sc starevol} code used extensively by 
Forestini, Siess, Charbonnel and collaborators. The first of these is the 2000 version
of {\sc starevol} \citep{SDF2000} as used by \citet{CZ07} and \citet{CL10}. We shall call this 
the {\sc starevol-geneva-v2.3} code, or {\sc se-g-v2.3} for short.
This code also uses the same opacity as the others above,
$\alpha_{MLT}=1.6$ and separate solution for the structure and the composition.
The results we quote for this code are taken from \cite{CL10}.

\subsection{\sc STAREVOL-V5.5}

This code was largely rewritten recently during development of the
binary stellar evolution code {\sc binstar} \citep{BINSTAR}.
In this version the nuclear burning and diffusive transport of the chemicals are
coupled and solved simultaneously, although it is of course possible
to disable that option and solve separately for the nucleosynthesis 
and then the mixing. We will use this feature later in some of our tests.
A major upgrade in the new version of this code is the
differencing of the equations, and the ability to simultaneously
solve for the structure variables as well as the composition 
(including both nuclear burning and mixing), as is done in {\sc stars}.
However this fully coupled solution
mode is slower and, as far as thermohaline mixing on the RGB is concerned,
tests showed that it does not alter significantly the evolution of the surface abundances.
Therefore we do not use this method in any of the tests described below.
Rather, we solve for the structure and then the coupled mixing and burning
(except in clearly identified tests where we uncouple the mixing and burning).
We shall refer to this code as {\sc se-v5.5.}

In both {\sc starevol} versions, the same criteria are used to
determine the evolution timestep $\delta t$. 
We employ the common sort of timestep constraints associated with changes in the structure variables.
We check that at every grid point the physical variables do not change by more than a specified 
value between consecutive models. We further limit the timestep during the pre-main sequence and RGB 
phases not to exceed a fraction of the Kelvin Helmholtz timescale $\tau_{KH}$. 
The expression we use is
\begin{equation}
t_{\rm KH} = {gM^2\over{RL}}.
\end{equation}
A more appropriate estimate would be to consider the envelope relaxation time scale 
\begin{equation}
t_{{\rm KH}_{\rm env}} = { {gM_{\rm env}M} \over{R_{\rm env}L}}
\end{equation}
but this does not make a significant difference. 
For low and
intermediate mass stars climbing the RGB, the constraint on
$\delta t$ is set ultimately by the requirement that the timestep does not exceed
some specified fraction of the Kelvin Helmholtz timescale ($\tau_{KH}$). In
{\sc se-v5.5}, this limit is set to  $\delta t < 0.3\tau_{KH}$.

\subsection{The Test Case}
Our test case will be a model of $1.25$\msun{} with ``solar'' composition, as defined in \citet{AGS05},
meaning X~$=0.7383$, Y=~$0.2495$ and Z~$=0.0122$. In particular we take the initial mass 
fraction of \chem{3}He to be $8\times 10^{-5}$ and the initial \chem{7}Li is given by A(Li)=3.25 on the ZAMS.
This is the case presented in some detail in \citet{CL10} and thus it allows us to compare the
results for that work, using {\sc se-g-v2.3}, with the other codes above, giving us a total of five 
essentially independent implementations of the thermohaline mixing algorithm to compare.

\begin{figure*}
\resizebox{0.7\hsize}{!}{\includegraphics{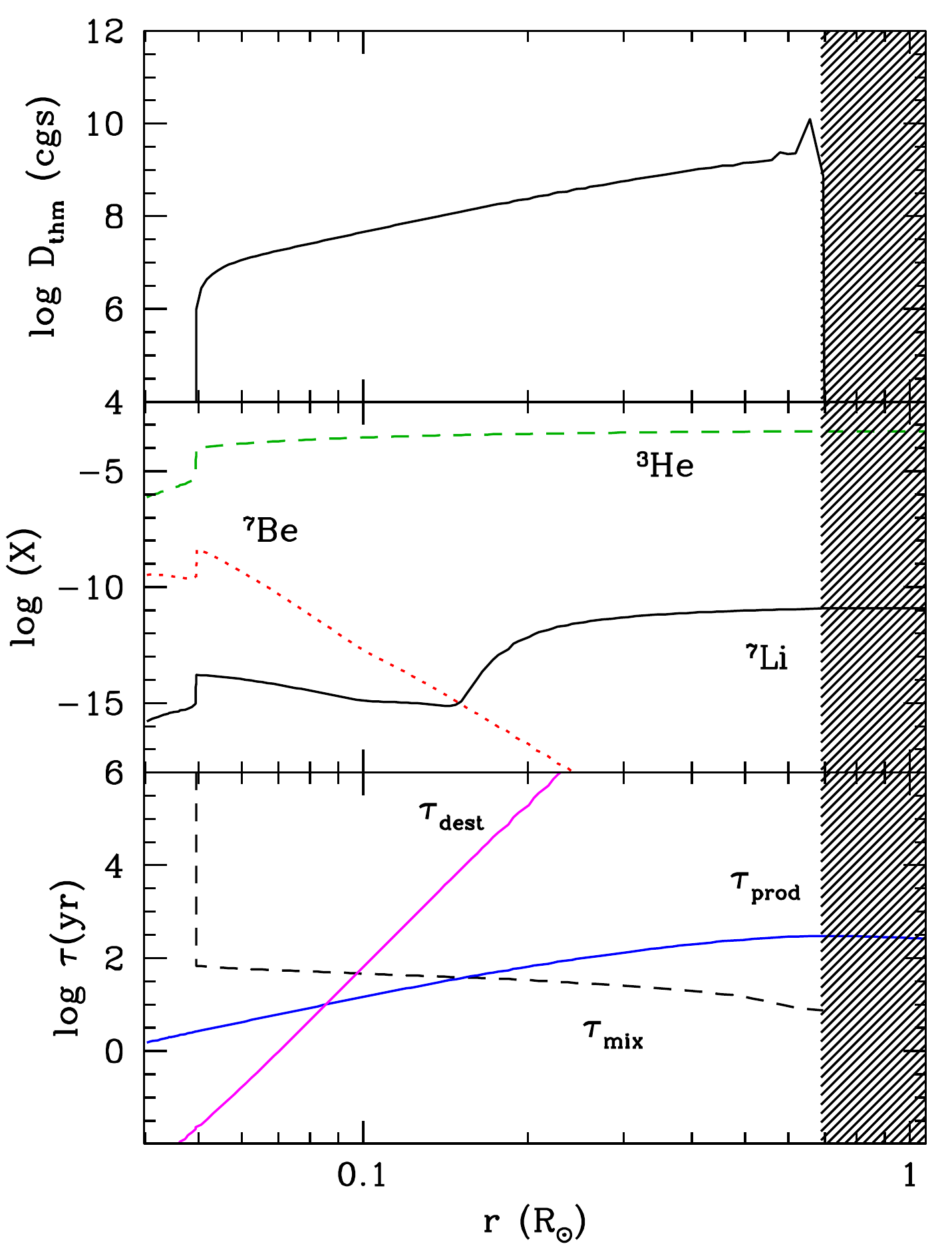}}
\caption{Internal structure in our test case soon after thermohaline mixing has begun,
when $L = 110\,$\Lsun, as calculated with {\sc se-v5.5}.
The plots cover the region from the location of the minimum in $\mu$ near the left hand edge
where $D_{\rm thm}$ goes to zero,  and the 
bottom of the convective envelope, shown by cross-hatching near the right hand edge. The top panel shows the
variation of $D_{\rm thm}$, the middle panel shows abundances of key species (in mass fraction), and 
important timescales are shown in the lower panel. Specifically these are the Li production timescale, i.e. the
timescale of the \chem{7}Be + $e^-$ reaction; the Li destruction timescale, being the timescale for
\chem{7}Li + p; and the timescale for mixing to transport material from the given position to the bottom of the
convective  envelope, defined by the sum of the local values $(\delta r)^2/D_{\rm thm}$ across the region,
where $\delta r$ is the width of the local mass shell.}
\label{InitialCase}
\end{figure*}

To introduce the important physics we show in Figure~\ref{InitialCase} the internal structure of our 
test model not long after thermohaline mixing has started, when the model 
is at $L=110\,$\lsun,  just above the
luminosity function (LF)  bump in the HR diagram. Because we will be concerned with mixing timescales we have
used radius on the $x$-axis instead of mass. For our purposes in this paper it is far
more intuitive to see the distances rather than masses. 
Note that in the deepest layers where the temperature is higher the shortest timescale is that for Li destruction.
Moving further out to cooler regions we
see that there is a small region where Li production dominates, and further out
again the mixing timescale becomes the shortest. Hence the effect of thermohaline mixing
is to feed material to the hot region where \chem{7}Li is rapidly destroyed. The destruction can be very quick,
with timescales as low as days, at the time shown in the Figure. Hence the important determinant for
the global evolution of the Li abundance is the timescale of feeding Li-rich material from the 
envelope into the hot region where it is almost instantly destroyed. Conversely, if Li (or any other species) 
is significantly produced then the important timescale is again that of the mixing which now drives the
increase in the envelope abundance. Hence it is vital  to resolve this mixing adequately.

\section{The Start of Thermohaline Mixing}
\begin{figure}
\resizebox{0.95\hsize}{!}{\includegraphics{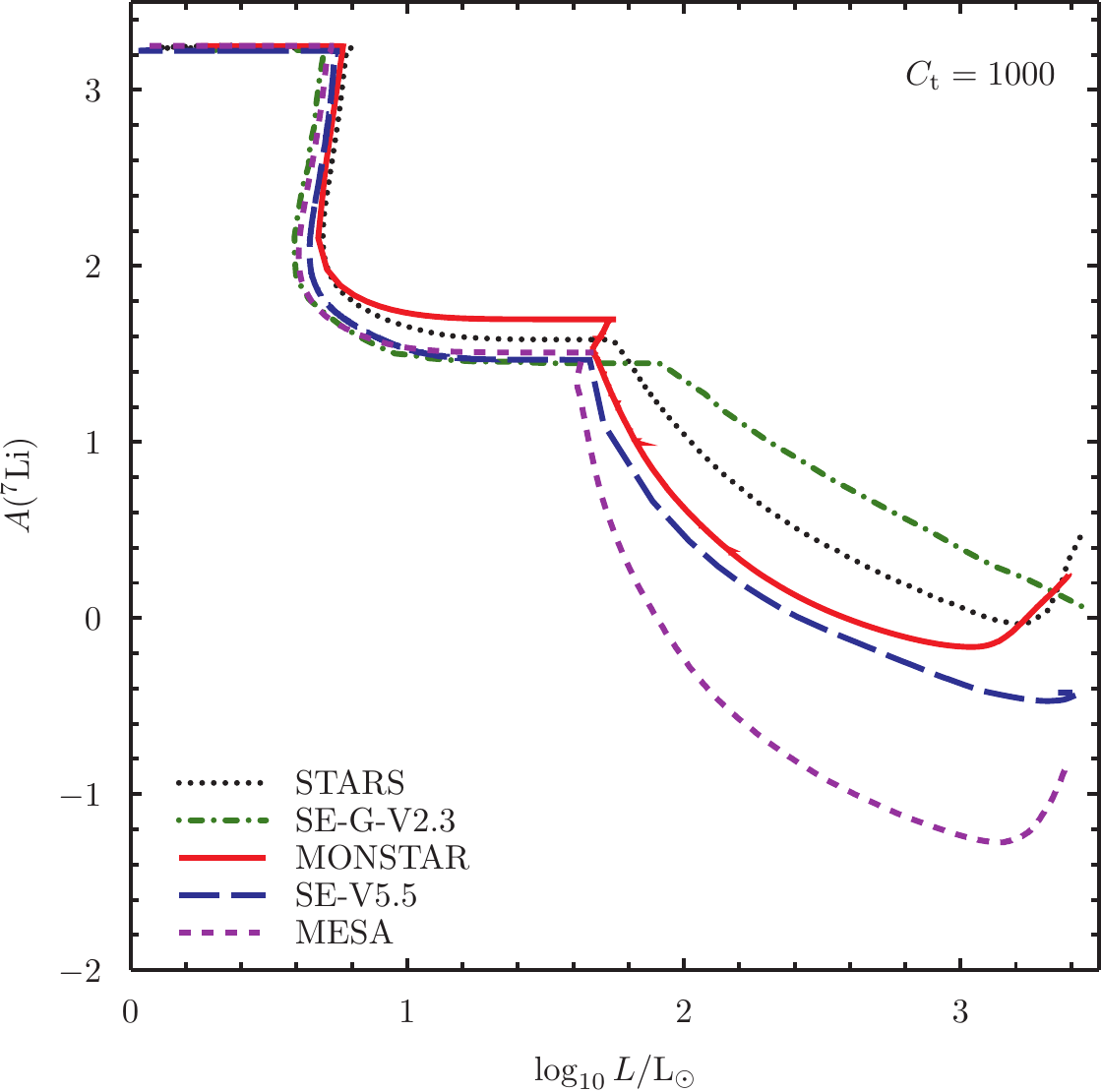}}
\caption{
Surface $A({\rm Li})$ plotted as a function of the luminosity $L$ 
from the various codes, all for our test
case model with $C_{\rm t}=1000$.  The figure shows the evolution from the main
sequence to the tip of the giant branch.  The lines show various codes run with
default resolution and timestepping.  The black dotted line is {\sc stars}, the
green dot-dash line {\sc se-g-v2.3}, the red solid line {\sc monstar}, the
blue long-dashed line {\sc se-v5.5} and the purple short-dashed line {\sc mesa}.
}
\label{Li1000}
\end{figure}

We start by showing in Figure~\ref{Li1000} the results from all of our codes for the case $C_t=1000$. 
It is pleasing to see that the plots all have a similar shape, showing the decrease 
in Li caused by FDU at log(L) $\simeq 0.7$, reaching a similar value A(Li) $\simeq 1.5$ until the start of 
thermohaline mixing, which occurs at log(L) $\simeq 1.7$--$2.0$ depending on the code. This then results 
in a further decrease of Li, except in some codes where there is a small {\it increase\/} found 
near the end of the evolution, at the tip of the giant branch.

The first point of difference concerns the first dredge-up. We define the depth of first dredge-up, 
$M_{FDU}$ as the maximum inward extent of the convective envelope. 
Clearly the final A(Li) value after FDU depends
on this value: simple conservation
of Li demands that the deeper the maximum extent of FDU the lower is A(Li).  
The luminosity at this time shall be denoted by $L_{FDU}$. When the hydrogen burning shell 
approaches the step in composition caused by the retreat of the convective envelope after FDU
the sudden increase in hydrogen fuel causes a hydrostatic readjustment of the structure, and
the radiated luminosity temporarily decreases before beginning to rise again with the new shell structure.
This produces the bump in the luminosity function seen in globular clusters.
Following \cite{CL10} we define the local minimum and maximum in 
luminosity as $L_{b,min}$ and $L_{b,max}$. These values are all independent of
thermohaline mixing. Table~\ref{FDU} gives these values for each of the codes used in this paper.
We note that the discrepancies between different codes at FDU is a topic of
considerable importance; rather than discuss the details here we refer the reader to a recent paper
on the subject by Angelou et al. (2014).

\begin{table*}
    \centering
    \caption{First Dredge-Up Parameters for our Calculations. For each of the codes used (listed in the first column)
             we give the minimum and maximum luminosities 
             at the bump, $L_{b,min}$ and $L_{b,max}$, respectively, as well as the
             deepest penetration of the convective envelope $M_{FDU}$ during first dredge-up,
             and the luminosity $L_{FDU}$ at that time. All values are in solar units.}
    \begin{tabular}{|lcccc|}
    \hline
    Code        & $L_{b,min}$/\lsun & $L_{b,max}$/\lsun & $M_{FDU}$/\msun & $L_{FDU}$/\lsun \\ \hline
    {\sc monstar}        & 47 & 57        & 0.272    & 25 \\
    {\sc stars}          & 49 & 52        & 0.262    & 27 \\
    {\sc mesa}           & 40 & 47        & 0.257    & 20 \\
    {\sc se-g-v2.3}      & 37 & 45        & 0.256    &--- \\
    {\sc se-v5.5}       & 42 & 50        & 0.262    & 20 \\ \hline
    \end{tabular}
\label{FDU}
\end{table*}

The next important difference is when the thermohaline mixing process starts to affect the surface composition. 
Obviously this depends on the maximum depth of the convective envelope during FDU, 
but there is also another effect at play.
For the surface Li abundance to show any change, we need the material at the bottom of the convective envelope
to be transported to where it is hot enough for Li to be destroyed. The thermohaline mixing
begins in the region where $\mu$ has decreased. But this does not initially extend all the way to the
bottom of the convective envelope. The region of the $\mu$ inversion  must develop, through mixing and burning,
so that there is a negative $\nabla_\mu$ from the minimum in $\mu$ to the bottom of the convective 
envelope. It is only then that the surface Li value will be seen to drop.  The luminosity at which this
happens is reported in Table~2 as $L_{thm}$ \citep[or $L_c$ in the notation of][]{CL10}.

\begin{table*}
    \centering
    \caption{The Start of Thermohaline Mixing. For each of the codes used (listed in the first column)
    we provide the value of $C_t$ used, the luminosity $L_{thm}$ (in solar units) when the surface composition
    shows the effects of thermohaline mixing (i.e. when the thermohaline region makes contact with the convective envelope), 
    and the time $\Delta t_{thm}$ elapsed between the beginning of thermohaline mixing and the 
    appearance of its effects at the surface (in units of $10^6$\,y). The third column lists the particular case
    being examined: ``norm'' refers to the typical values (for that code) for space and time resolutions;
    changes to the standard time-step criteria are given as factors of $\delta t$ or restrictions
    in terms of the Kelvin-Helmholtz timescale $\tau_{KH}$;  mesh-spacing criteria are detailed in the text
    or given as the total number of $N$ of zones. The case $\dot{M}=0$
    had mass-loss turned off, but the usual space and time resolutions. {The notation $\mu$
    indicates cases for {\sc stars} where the calculation of $\mu$ was made using the current composition
    and not that of the previous timestep (which is the usual case for that code).
    }}
    \begin{tabular}{|llccc|}
    \hline
    Code            & $C_t$   &Case   & $L_{thm}$/\lsun & $\Delta t_{thm}$ ($10^6$ y)  \\ \hline
    {\sc monstar}   & $100$   &norm   & 54          & 3.85    \\
    {\sc stars}     & $100$   &norm   & 59          & 5.44    \\
    {\sc mesa}      & $100$   &norm   & 42          & 5.96  \\
    {\sc se-v5.5}   & $100$   &norm   & 47          & 2.63    \\
\hline
    {\sc monstar}   & $1000$  &norm   & 53          & 3.24    \\
    {\sc stars}     & $1000$  &norm   & 59          & 5.44    \\
    {\sc mesa}      & $1000$  &norm   & 43          & 5.30  \\
    {\sc se-g-v2.3} & $1000$  &norm   & 94          & 44.6    \\
    {\sc se-v5.5}  & $1000$  &norm   & 47          & 2.19    \\
\hline
    {\sc monstar}   & $10\,000$ &norm   & 55          & 2.99    \\
    {\sc stars}     & $10\,000$ &norm   & 59          & 5.48    \\
    {\sc mesa}      & $10\,000$ &norm   & 46          & 5.64  \\
    {\sc se-g-v2.3} & $10\,000$ &norm   & 94          & ---     \\
    {\sc se-v5.5}   & $10\,000$ &norm   & 47          & 2.07    \\
\hline
    {\sc monstar}   & $1000$  &$\delta t*100$           & 54   & 3.44    \\
    {\sc monstar}   & $1000$  &$\delta t*1.0$\footnote{This is the ``norm'' case.}         & 55   & 3.24    \\
    {\sc stars}     & $1000$  &$\delta t*1.0^a$         & 59   & 5.44    \\
    {\sc stars}     & $1000$  &$\delta t*0.1$           & 52   & 0.77    \\
    {\sc stars}     & $1000$  &$N=499$            & 57   & 1.21    \\
    {\sc stars}     & $1000$  &$N=1999$           & 58   & 5.31    \\
    {\sc stars}     & $1000$  &$\delta t, \mu$                & 52   & 0.91    \\
    {\sc stars}     & $1000$  &$\delta t*0.1, \mu$            & 52   & 0.99    \\
    {\sc se-v5.5}  & $1000$  &$\dot{M}=0$        & 45   & 2.39    \\
    {\sc se-v5.5}  & $1000$  &$\delta m$ low res         & 45   & 1.57    \\
    {\sc se-v5.5}  & $1000$  &$\delta m$ norm$^a$        & 47   & 2.19    \\
    {\sc se-v5.5}  & $1000$  &$\delta m$ high res        & 46   & 3.31    \\
    {\sc se-v5.5}  & $1000$  &$\delta t \la 1.0\tau_{KH}$ & 44   & 9.57    \\
    {\sc se-v5.5}  & $1000$  &$\delta t \la 0.3\tau_{KH}^a$   & 47   & 2.19    \\
    {\sc se-v5.5}  & $1000$  &$\delta t \la 0.1\tau_{KH}$ & 42   & 1.65    \\ \hline
    \end{tabular}
\label{startthmtab}
\end{table*}

\begin{figure*}
\includegraphics[width=0.48\textwidth]{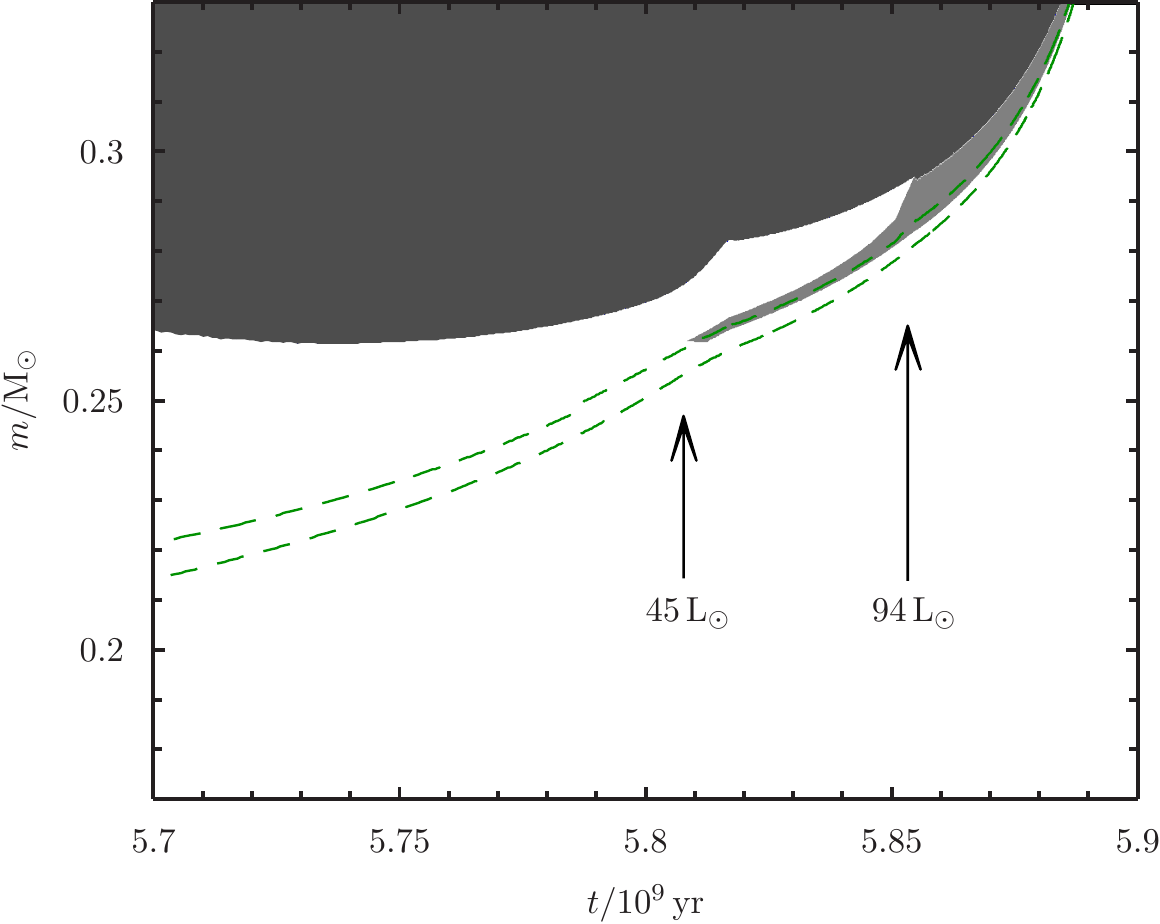}
\includegraphics[width=0.49\textwidth]{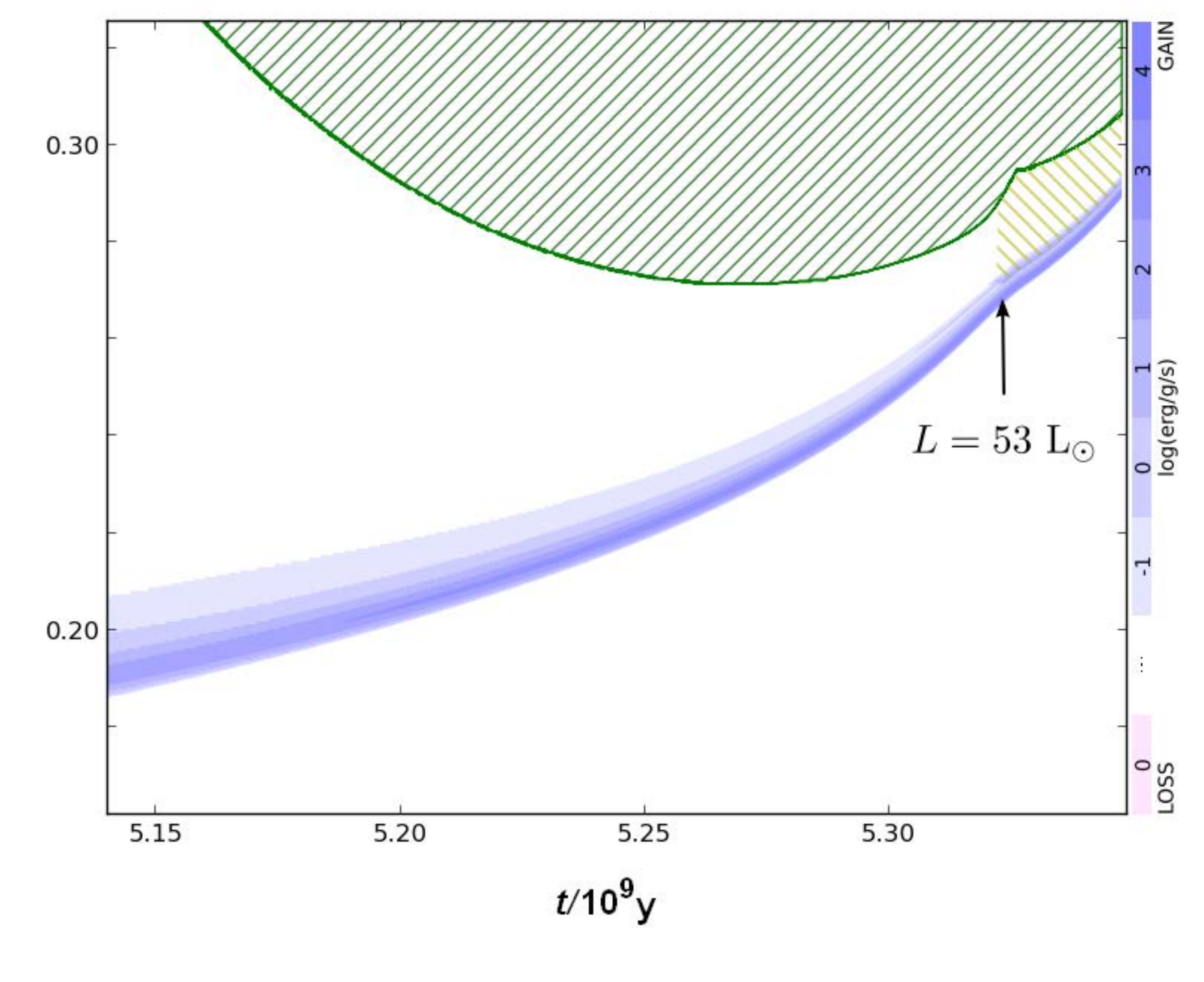}
\caption{The start of thermohaline mixing in {\sc se-g-v2.3} (left panel) and
{\sc monstar} (right panel), with $C_t=1000$.  Shaded regions show the locations
in mass of convective mixing (dark shading) and thermohaline mixing (light
shading) as a function of time since the zero-age main sequence.  Also marked
are the boundaries of the hydrogen-burning shell (green dashed line for {\sc
se-g-v2.3}; blue shaded region for {\sc monstar}).  Also shown are the
luminosities when thermohaline mixing begins and when the mixed region connects
to the envelope (these are effectively the same  in {\sc monstar}).}  
\label{startthm1}
\end{figure*}

So just how long does it take for the 
thermohaline region to make contact with the convective envelope?  Figure~\ref{startthm1} shows the results for
two cases: {\sc se-g-v2.3} \citep{CL10} and  {\sc monstar}.  The time 
delay between the start of thermohaline mixing and the appearance of its effects 
at the surface is given in Table~\ref{startthmtab} as $\Delta t_{thm}$.
In {\sc se-g-v2.3} the connection between the two mixing zones takes about 45 million years, 
whereas in {\sc monstar} it occurs in about 3 million years.
The development of this region is crucially dependent on the propagation 
of the $\mu$ inversion and requires an accurate determination of the mixing because it is the mixing which feeds
the burning and produces the inversion. Fresh material rich in \chem{3}He must be brought down to the burning region, 
where it is transmuted into material with a lower $\mu$ prior to being transported away from the burning region. 
If the timestep is too large then we do not follow this development accurately with the result that the 
thermohaline mixing takes longer to connect to the convective envelope and hence for the surface abundances to 
show the effect of the burning.

\begin{figure*}

\begin{center}
\includegraphics[width=0.49\textwidth]{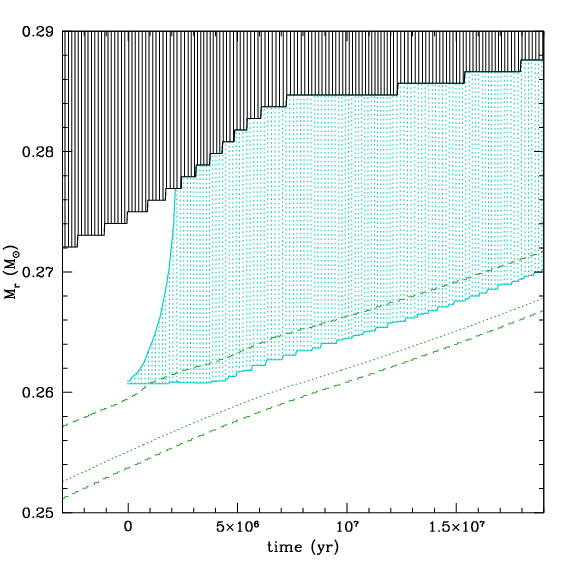}
\includegraphics[width=0.49\textwidth]{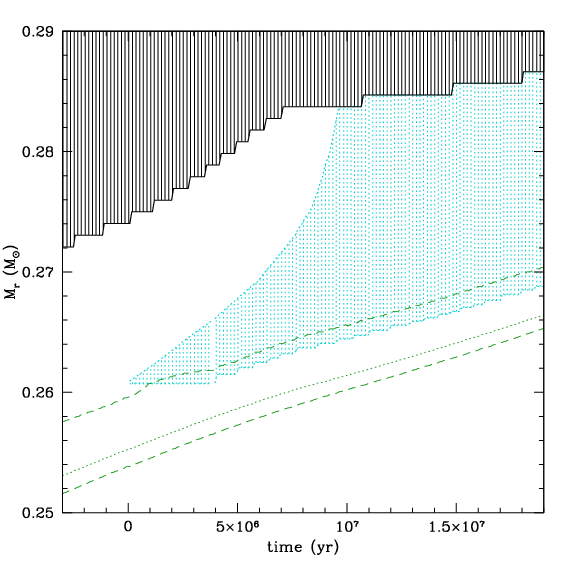}
\end{center}
\caption{The start of thermohaline mixing in {\sc se-v5.5} with timesteps limited to 
$0.3\tau_{KH}$ (left panel) and $\tau_{KH}$ (right panel). Time is set to zero
when thermohaline mixing first appears. The hydrogen burning shell is present
between the two dashed lines and the location of its maximum energy output is
shown by the dotted line.  The blue shaded region shows the extent of
thermohaline mixing, the grey shaded region the convective envelope.
}
\label{startthm2}
\end{figure*}

Note that the mixing is driven by a $\mu$ inversion that is as small as 1 part in $10^6$ or so initially. 
Indeed, as soon as  any decrease in $\mu$ is resolved, then the algorithm will start mixing. 
How this propagates will depend on how the $\mu$ inversion is resolved and one can understand why
small timesteps are needed. One can also imagine that in a real star, with magnetic fields, 
rotation, gravity waves etc, such small variations may not be established as easily as in this idealization
we are modelling \citep{MMLC2013}. We remind the reader that we are concerned
here with an accurate calculation
within the paradigm of the theory, rather than a test of the validity of the theory itself.

In any event we tried some tests with {\sc se-v5.5}. We took large timesteps, $\delta t,$ limited to be less than 
the total stellar Kelvin-Helmholtz timescale, $\tau_{KH}$, and small timesteps limited to $0.3\tau_{KH}$.
The differences are seen in Figure~\ref{startthm2}, with $\Delta t_{thm} \simeq 9.57$ million 
years for the larger time 
steps but reduced to only 2.19 million years when the smaller timesteps were used; we discuss the reasons
for this below.

The next test was also performed with {\sc se-v5.5}.
The Kelvin-Helmholtz timescale for the model is 68\,000y at this stage of the evolution\footnote{We note that this timescale becomes much shorter later in the evolution on the RGB.}.
Hence  we used fixed $\delta t$ of 68\,000y for the structure but took ten nucleosynthesis
sub-steps where we kept 
the structure constant but mixed and burnt the composition. We did not update $D_{thm}$ or $\nabla_\mu$, just the composition. 
This produced a slower growth of the thermohaline region: in one million years the region grows by only $0.002$\msun,
whereas the convective envelope is about $0.013$\msun{} from the 
region where the thermohaline mixing begins (as shown in Figure~\ref{startthm2}).
In the next test we also recomputed the $\nabla_\mu$ in each of the sub-steps and hence updated the 
diffusion coefficient. That resulted in more rapid growth, with $\Delta t_{thm} \simeq 1.9$ million years. 
This shows clearly that the feedback from the mixing on the diffusion coefficient is crucial for an 
accurate solution, and hence small timesteps must be taken.

The thermohaline mixing timescale over a distance $\delta r$ is just $\delta r^2/D_{thm}$. So from the size of the mesh 
in the calculation at this stage we estimate the mixing time to be about 1000 years between mass shells. 
The thermohaline
region was resolved with about 50 mass zones. Hence material could be mixed over just a few zones (about 7) 
in each of the 6\,800y sub-steps.
We then repeated the calculation with  single evolutionary time steps of 3400 years. 
The mixing and burning is calculated without sub-steps each time the structure is converged.
Here the material should diffuse no more than 3 or 4 zones per timestep. This should  resolve
the behaviour much better, and it gives a total time for the thermohaline region to reach the convective 
envelope of about 1.5 million years, which agrees with the other codes when using small $\delta t$, as 
well as {\sc se-v5.5}
results with $\delta t$ limited to be $0.1\tau_{KH}$, as reported in Table~\ref{startthmtab}.

As a final test we tried reverting to large timesteps (68\,000y) but updating $\nabla_\mu$ and $D_{thm}$ after each 
iteration in the nucleosynthesis routine. 
In a run of 1.5 million years the thermohaline region had only reached about 1/3 of the way to the
convective envelope.
We conclude that there is no alternative but to take small timesteps during this part of 
the evolution as it determines when the surface abundance will begin to change. It is crucial that the 
value of $D_{thm}$ be updated after even small changes in the composition  -- this drives the mixing and 
hence feeds back on itself. We are reasonably confident that the {\it correct\/} value for 
$\Delta t_{thm}$ (i.e. within the 1D diffusive 
model of \citet{Ketal80}  and \citet{Ulrich72} and for this particular stellar model) is about 1 million years.
This is significantly different to the \citet{CL10} value of 45 million years, which presumably did not resolve
this initial phase adequately. It is this delay that is responsible for the offset in the {\sc se-g-v2.3} results for
Li destruction, as shown in Figure~\ref{Li1000}. This is also why the value of $L_{thm}$ is much larger (at $94$\lsun) 
than the other codes tested (at about 50--60\lsun; see Table~\ref{startthmtab} for a list of the tests performed).

Finally, we should be able to estimate the timescale
for the growth of the thermohaline region from the structure. We have an estimate of the
timescale of a diffusion process to move matter over a distance $\delta r$ as $\tau \simeq (\delta r)^2/D_{thm}$.
Note that this is the time for material {\em in the thermohaline region\/} to move a distance $\delta r$.
One may initially think that this is not the rate at which the borders of the zone grow. For example,
the convective turnover time in a convective envelope is in no way related to the time it takes for
first dredge-up to reach its deepest extent. The material inside the convective envelope
moves with one speed, quite independent
of the rate at which the edge of the convective zone moves. In this case the inner edge of the
convective envelope is determined by the stellar structure responding to the expansion of the star
as it ascends the giant branch, and is quite independent of the concomitant changes in abundance. 
The speed with which the inner edge of the convective envelope moves inward is typically 
$v_{FDU} \simeq 10^{-5}$cm/sec,
quite different to the speed of matter in the convective envelope, which the mixing length theory gives
as $v_{conv} \simeq 10^4$cm/sec. In the case of diffusion, however, the two speeds are the same. The edge
of the mixed region moves {\em precisely\/} because it is diffusing into the homogeneous region. Thus the growth of the
thermohaline mixing region can be estimated from $D_{thm}$ in the region. When thermohaline mixing first appears,
long before the region has made contact with the convective envelope, we have $D_{thm} \simeq 10^6 - 10^7$cm$^2$/s.
Later when the regions have nearly joined $D_{thm}$ has grown to $10^9$cm$^2$/s. Given that the
mixing must cross a region of about $0.5$\rsun{} we estimate that at the start the initial
timescale for reaching the envelope is $\tau \simeq (\delta r)^2/D_{thm} \simeq 3 - 30$Myr. But the diffusion coefficient
increases rapidly and by the time the regions have joined the timescale is down to $\tau \simeq 0.03$Myr. This
can only give us a crude consistency check, but it
supports the result of our calculations giving  about 1 million years for the actual time.

We also performed some tests where we varied the spatial mesh, and these are reported in
Table~\ref{startthmtab}. We found that the mesh spacing was not nearly as critical as the
timestep, a result which will re-appear below.

\section{The Evolution of the Surface Abundance of \chem{7}Li}

\begin{figure}
\resizebox{0.49\textwidth}{!}{\includegraphics{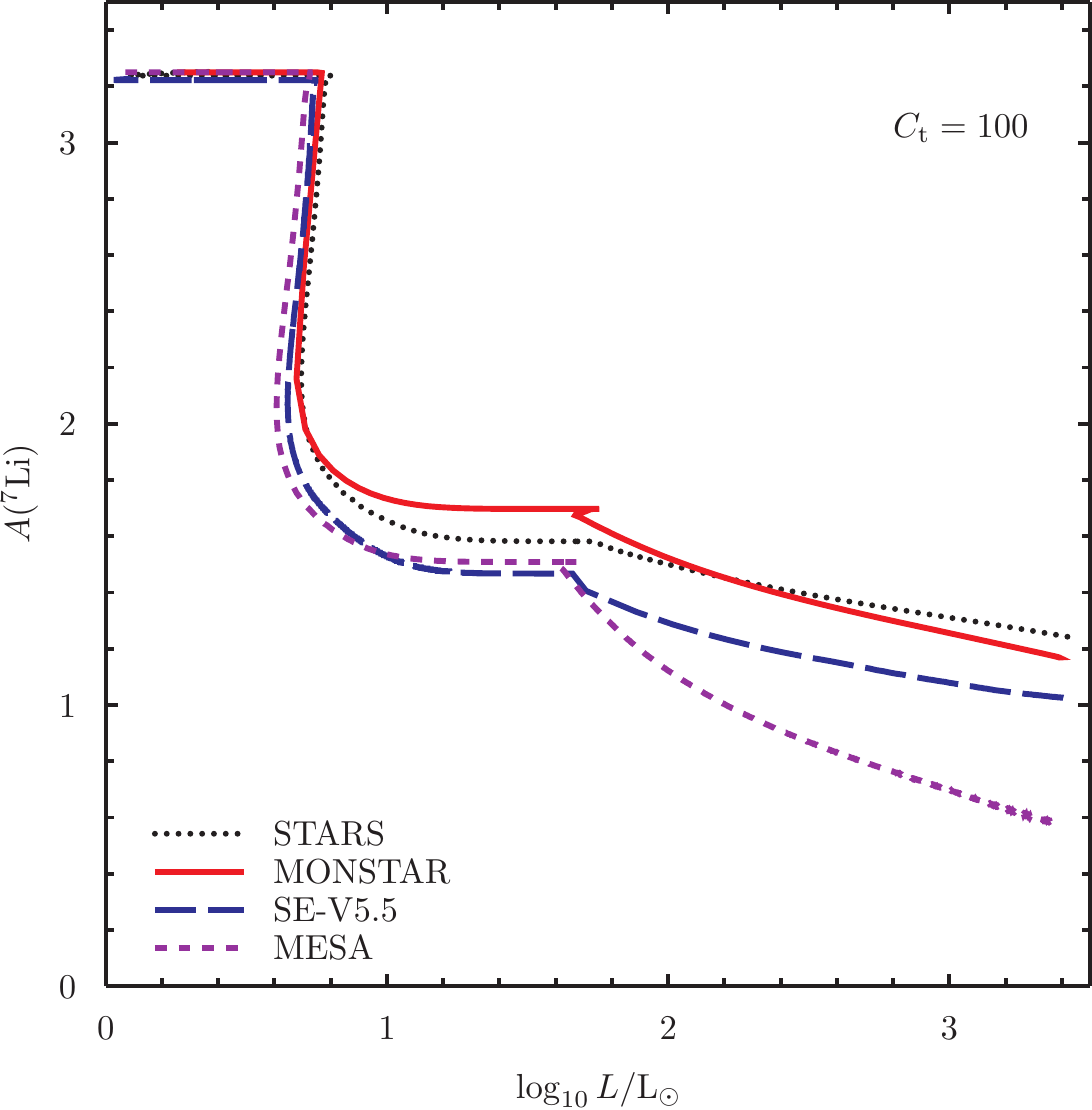}}
\caption{
Surface $A({\rm Li})$ plotted as a function of the luminosity $L$ for our test
case model with $C_{\rm t}=100$.  The figure shows the evolution from the main
sequence to the tip of the giant branch.  The lines show various codes run with
default resolution and timestepping.  The black dotted line is {\sc stars}, the
red solid line {\sc monstar}, the
blue long-dashed line {\sc se-v5.5} and the purple short-dashed line {\sc mesa}.
}
\label{Li100}
\end{figure}

We show the results for $C_t=100$ in Figure~\ref{Li100}. Note that we do not have results for 
{\sc se-g-v2.3} for this case.
There is some scatter in the results, and by the time the models reach the tip of the giant 
branch there is 1 dex spread in the predicted final values for the Li content. 
This is clearly not acceptable, but as we will see, 
the situation gets worse as $C_t$ increases.

\begin{figure}
\resizebox{0.48\textwidth}{!}{\includegraphics{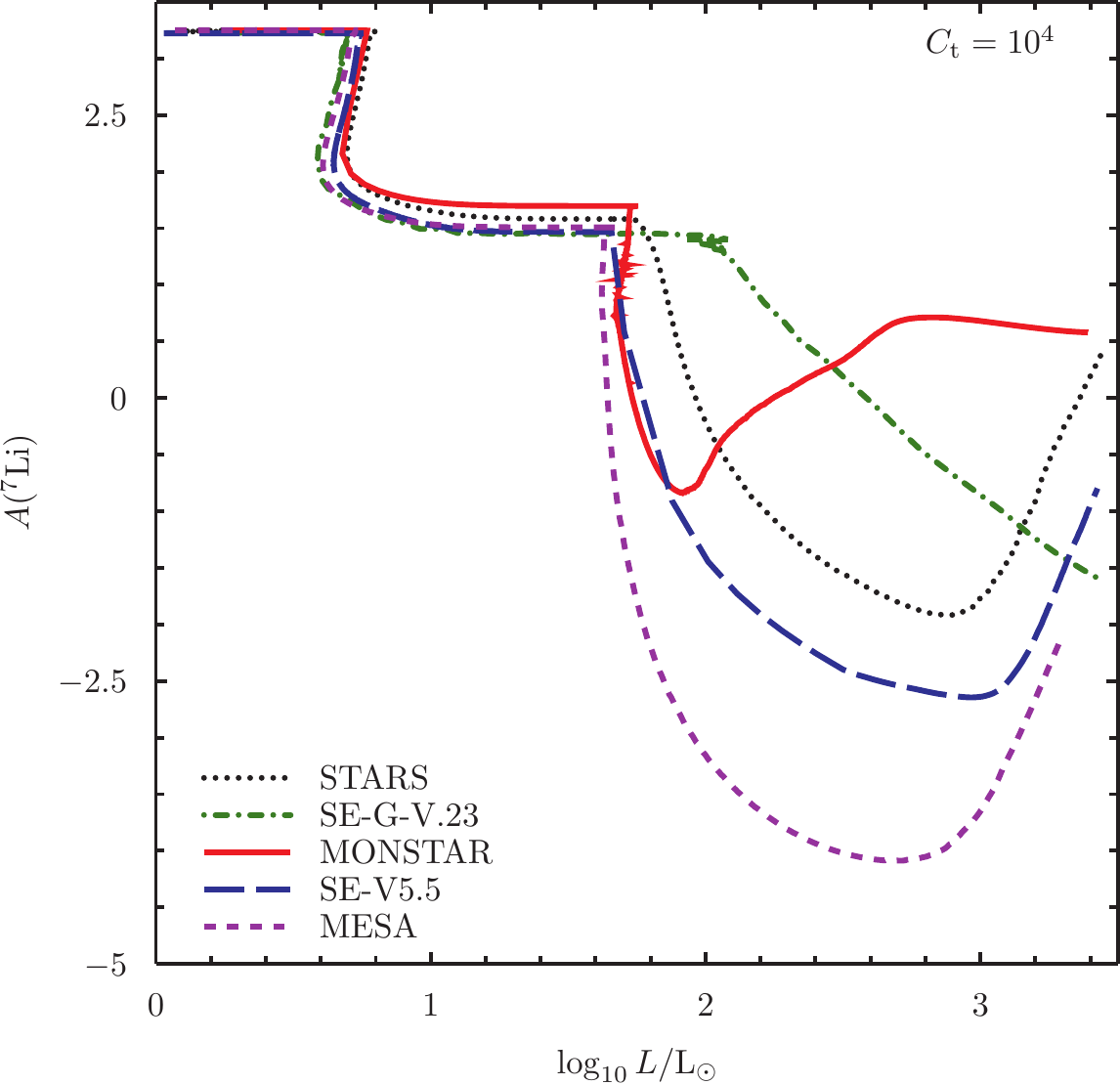}}
\caption{
Surface $A({\rm Li})$ plotted as a function of the luminosity $L$ for our test
case model with $C_{\rm t}=10\,000$.  The figure shows the evolution from the main
sequence to the tip of the giant branch.  The lines show various codes run with
default resolution and timestepping.  The black dotted line is {\sc stars}, the
green dot-dash line {\sc se-g-v2.3}, the red solid line {\sc monstar}, the
blue long-dashed line {\sc se-v5.5} and the purple short-dashed line {\sc mesa}.
}
\label{Li10000}
\end{figure}

We have already shown in Figure~\ref{Li1000} the cases with $C_t=1000$, including {\sc se-g-v2.3} this time, from \citet{CL10}.
The early evolution is the same in all codes, but once the thermohaline mixing starts we get a very diverse spread of results.
There is a spread of up to 2 dex in the Li abundance from different codes, and we even see
an increase in the surface A(Li) near the tip of the giant branch in some cases, but not all. When we move to $C_t=10\,000$
(Fig~\ref{Li10000}) then all codes except {\sc se-g-v2.3} show the increase of Li near the tip of the RGB,
and a maximum spread in A(Li) between codes of over 3 dex.
To try to determine the cause of this variation we have performed a number of tests with the different codes.

\subsection{Testing Temporal Resolution}
\begin{figure}
\resizebox{0.48\textwidth}{!}{\includegraphics{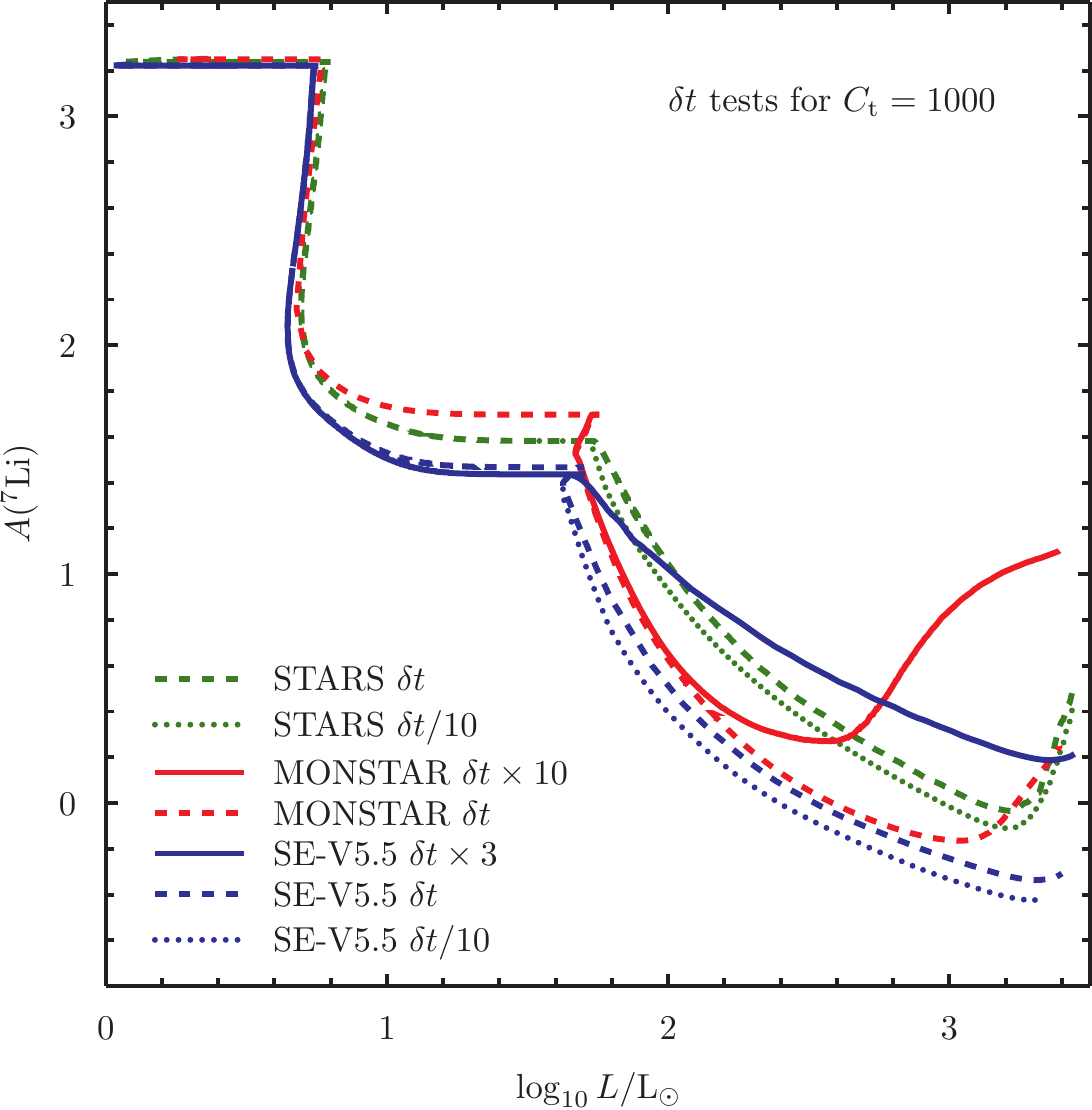}}
\caption{
The effects of varying the timestep on surface $A({\rm Li})$ as a function of
luminosity $L$ for our test case model with $C_{\rm t}=1000$.  The figure shows
the evolution from the main sequence to the tip of the giant branch.  The green
lines
are {\sc stars} run with default timestep (dashes) and timestep reduced by a
factor of 10 (dots).  The  
red lines are  {\sc monstar} with default timestep (dashes) and timestep
increased by a factor of ten (solid).  Blue lines are {\sc se-v5.5} with the
default timestep (dashes), the timestep increased by a factor of three (solid
line) and reduced by a factor of ten (dotted line).
}
\label{dttests}
\end{figure}

We begin by testing the effect of the timestep, since we have seen how crucial this is in the early stages of the evolution.
We chose the case with $C_t=1000$, which is the preferred value for
explaining various observations, as discussed above.
Tests were run with three codes, using the ``normal'' timestep 
constraints,  which we refer to as ``$\delta t$'', 
and then with increased or decreased timesteps given as multiples of 
this standard $\delta t$. The results are shown in Figure~\ref{dttests}.

Firstly consider the runs done with {\sc se-v5.5}. The ``normal'' case and the case with
timesteps reduced by a factor of 3 seem to give essentially the same results. 
Contrarily, if the timestep is increased by a factor of 3 then the results start to differ, with 
less Li being destroyed due to the code not being able to follow accurately the flow of material 
into the top of the hydrogen burning shell where Li is efficiently destroyed. 

Now consider the {\sc monstar} runs. Here the standard case seems to agree well with the converged cases
found with the {\sc se-v5.5} code. However if the timestep is increased by a factor of 10 then the code is able
to match the Li destruction initially, but it starts to depart from the (presumably) correct 
solution and indeed begins to produce Li as the model nears the tip of the RGB. Note that Li production is
seen in the normal case also, for this code, but the production is not as significant with the smaller timestep.
Upon closer examination there is also a hint of an increase in Li in the well resolved calculations by 
the {\sc se-v5.5} code. We address the cause of this rise in Li in \S\ref{secLirise}.

Two tests were run with the {\sc stars} code also, as shown in Figure~\ref{dttests}. 
This code does not destroy as much Li as
{\sc monstar} and {\sc se-v5.5}, and it has not yet converged 
on our inferred solution. We note that it does show the same trends 
as the other codes, in that increasing the timestep
stops the calculation from adequately following the flow of material from the convective
envelope into the hot burning region resulting is a
lower Li depletion, as will be explained in Sect 5.3.

As stated earlier, {\sc stars} normally uses the composition at the previous timestep
to calculate $\mu$. This is not ideal for the present case where there is 
strongly coupled feedback between the mixing and the composition. To test this we
ran cases where $\mu$ was calculated using the current composition. These are reported
in Table~2 with the notation ``$\mu$'' and either the normal timestep criteria, $\delta t$,
or 1/10 of those criteria, $\delta t*0.1$. In practice, the timesteps had to be
greatly reduced to ensure convergence, and the results are very similar to the standard case
with $\delta t*0.1$.

\subsection{Testing Spatial Resolution}

\begin{figure}
\resizebox{0.48\textwidth}{!}{\includegraphics{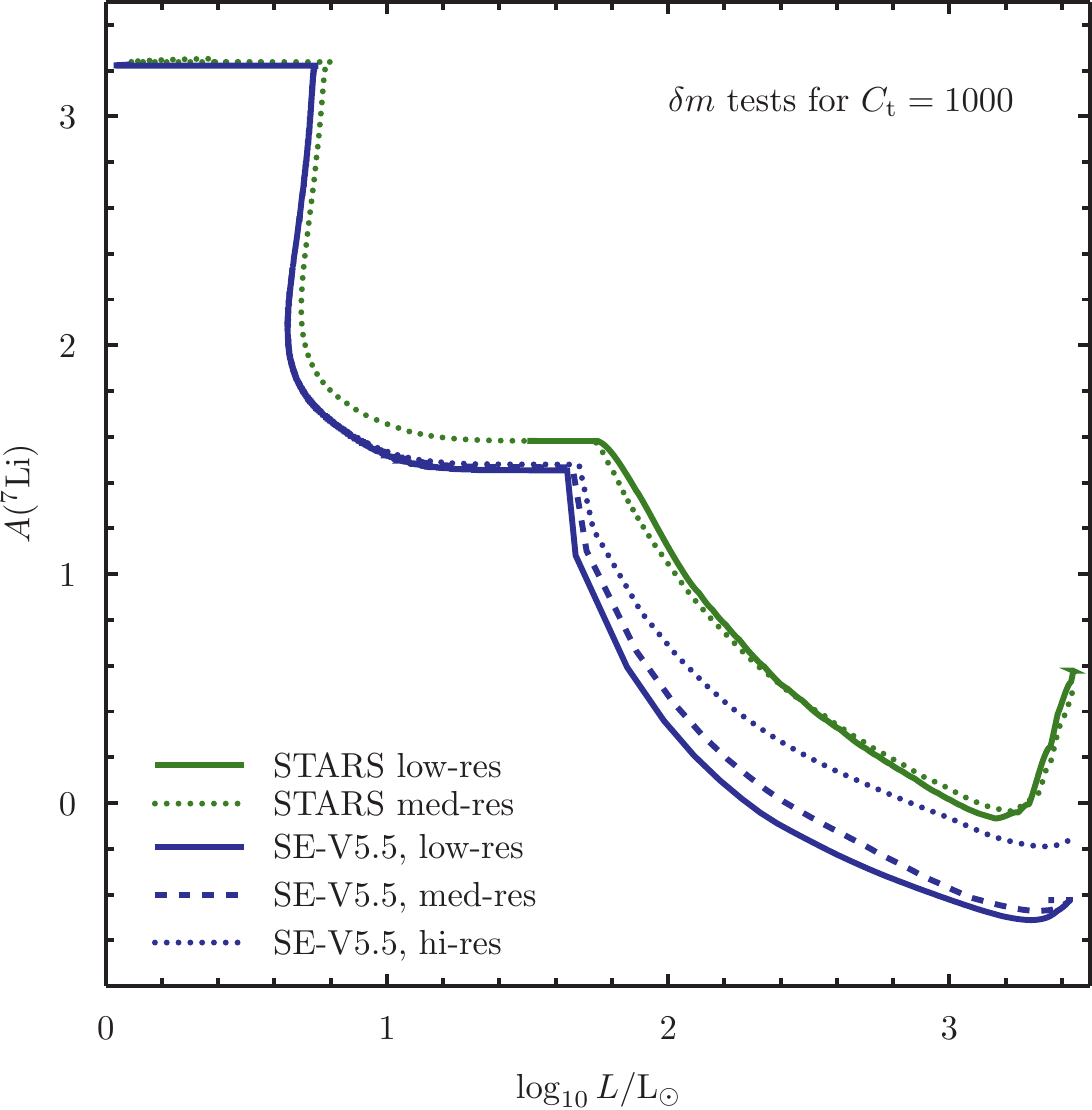}}
\caption{
The effects of varying the spatial resolution on surface $A({\rm Li})$ as a
function of luminosity $L$ for our test case model with $C_{\rm t}=1000$.  The
figure shows the evolution from the main sequence to the tip of the giant
branch.  The green lines
are {\sc stars} runs with low resolution (499 mesh points, solid line) and medium
resolution (1999 mesh points, dotted line). Blue lines are {\sc se-v5.5} with the
default resolution (100 zones in the thermohaline region, dashed line), low
resolution (60 zones in the thermohaline region, solid line) and high resolution
(170 zones in the thermohaline region, dotted line).
}
\label{dmtests}
\end{figure}

We performed tests on the spatial resolution using {\sc stars} and {\sc se-v5.5}. Specifically,
for {\sc stars}  we ran one case with 1999
mesh points, called med-res, and one case with 499 mesh points, called lo-res. As shown in
Figure~\ref{dmtests} as we increase the spatial resolution we actually burn less Li. This is in the opposite
direction to the case where we increase the time resolution, and the reason for this will be discussed below.
Similar results were found for the {\sc se-v5.5} code in the three cases presented. In this case the low, standard and
high resolution cases are defined in terms of the number of zones in the thermohaline region, being 60, 100
and 170 respectively. Again, increasing the resolution shows that less Li is burned. Both codes agree in this respect, although 
again the {\sc stars} code seems to be less efficient in destroying Li than the other codes.

\subsection{Understanding the Resolution Tests}

Consider solving the diffusion equation for a quantity $u$
\begin{equation}
\frac{\partial u}{\partial t}
   = D \frac{\partial^2 u}{\partial x^2}
\end{equation}
over a region with $N$ zones of constant width $\delta x$ and with constant timestep $\delta t$
and constant diffusion coefficient $D$.
We can write this as a fully implicit scheme as:

\begin{equation}
\frac{u^{n+1}_j-u^{n}_j}{\delta t} = D 
   \left(
   \frac{u_{j+1}^{n+1} - 2u_j^{n+1} + u_{j-1}^{n+1}}{(\delta x)^2}    
   \right)
\end{equation}
where superscripts refer to the timestep and subscripts refer to the spatial zone.
We assume the boundary conditions $u_0$ = constant and $u_N$ = constant.
Hence can write this as the matrix equation
\begin{equation}
{\bf u}^{n} = {\bf M} {\bf u}^{n+1}
\end{equation}
where 
\begin{equation}{\bf u}^n = (u_0^n, u_1^n, \ldots , u_N^n)^T\end{equation}
and
\begin{equation}
M = 
\left(
\begin{array}{ccccccc}
1        & 0            & 0            & 0            & 0       &\cdots &0 \\
-\alpha  & (1+2\alpha)  & -\alpha      & 0            & 0       &\cdots &0 \\
0        &-\alpha       & (1+2\alpha)  & -\alpha      & 0       &\cdots &0 \\
0        &0             &-\alpha       & (1+2\alpha)  & -\alpha &\cdots &0 \\
\vdots   &\vdots        &\vdots        &\vdots        &\vdots   &\ddots &\vdots \\
0        &0             &0             &0             &0        &\cdots &1 \\
\end{array}
\right).
\end{equation}

Here the parameter $\alpha$ is simply

\begin{equation}
\alpha = D\delta t/(\delta x)^2.
\end{equation}

This is a tri-diagonal system which can be solved easily
by the Thomas Algorithm, for example. It is always stable because the diagonal elements are
always larger than the sum of the magnitudes of the sub- and super-diagonal elements. i.e.
\begin{equation}
|1 + 2\alpha| > |\alpha| + |\alpha|
\end{equation} 
for all $\alpha > 0$, which is the case here.

The salient feature here is that, for this case with constant space and time steps, the 
solution is determined entirely by $\alpha$, which enables us to relate the chosen spatial and temporal
mesh to the solution. For example, halving the timestep $\delta t$ reduces $\alpha$ by a factor of two.
This is the same as using the original timestep but with $D$ reduced by a factor of two. Hence we
expect the solution with finer timesteps to be similar to solving the original equation with a smaller $D$.
Similarly, if we halve the spatial mesh $\delta x$ then $\alpha$ increases by a factor of 4, so that is 
equivalent to the using the initial $\delta x$ but an increase in $D$ by a factor of 4. 

Thus increasing the resolution in space and in time have opposite effects on the solution, which  is
consistent with the tests performed in the previous section. Of course, these changes are only 
indicative; for instance, when halving the timestep we should actually compare the solution 
after two time steps with the original solution with the single timestep.  Similarly
when we double the number of mass shells we should compare every second mesh-point in 
the new solution with the original solution. Nevertheless, the discussion here gives some
explanation for why the solution behaves differently for changes in $\delta x$ and $\delta t$.

\section{Determining Appropriate Resolutions for an Accurate Solution}
\subsection{Timesteps}
Understanding the dependence of the solutions on the time and mesh spacing is 
important but it does not guide us in finding an appropriate discretization for our calculations.
We now address this question, through an examination of Figure~\ref{InitialCase}.
The lowest panel in the figure shows the important timescales. For the majority of the evolution
the Li is destroyed through mixing to the hot region at the bottom of the thermohaline zone.
Hence it is crucial to determine the rate at which this mixing occurs. In the region where destruction dominates,
this nuclear burning is very fast. The efficiency of the destruction thus depends on how rapidly it can be fed by the mixing.
Likewise, in the case where production dominates, which we discuss below, the rate of increase is determined by the
rate at which the created Li can be injected into the envelope. 

Thus the region where the Li production and destruction timescales are equal is going to be
crucial in determining the evolution of the surface Li content. 
Let us call this the equilibrium point. We must adequately resolve the region between this point and the
convective envelope if we are to follow the Li transport accurately. If the timestep is too long,
or the spatial mesh too large, we connect the two regions at a different rate to what they would achieve in reality.
We will define the timescale for mixing from this point to the bottom of the convective envelope as $\tau_0$,
and we suggest this is used as a typical timestep for the calculation. 

To calculate this $\tau_0$ we note that $D \simeq v l$ where $v$ is the local mixing velocity and $l$ is the
mean free path, or mixing length, of the moving parcel of gas. Hence at mass shell $i$ we have
\begin{equation}
v_i = \frac{D_i}{l_i}.
\end{equation}

Let $\delta r_i$ be the radial width of each mass shell.
Then the mixing timescale from shell $i$ to the convective envelope is
\begin{equation}
\tau_0 = \sum_i \frac{\delta r_i}{v_i} = \sum_i \frac{\delta r_i l_i}{D_i}
\end{equation}
\noindent where the sum extends from the current mesh point to the bottom of the convective envelope.
If we write $l_i = a_i \delta r_i$ 
for some real numbers $a_i$ 
then we get
\begin{equation}
\tau_0 = \sum_i \frac{a_i (\delta r_i)^2}{D_i} \propto \sum_i \frac{(\delta r_i)^2}{D_i}.
\end{equation}
Hence we will define our preferred timestep $\delta t_0$ by
\begin{equation}
\delta t_0 \equiv \tau_0 = \sum_i \frac{(\delta r_i)^2}{D_i}
\label{eqn:deltat0}
\end{equation} with the knowledge that we have
ignored some constants, but maintained the essence of the physics. 

\begin{figure}
\resizebox{0.48\textwidth}{!}{\includegraphics{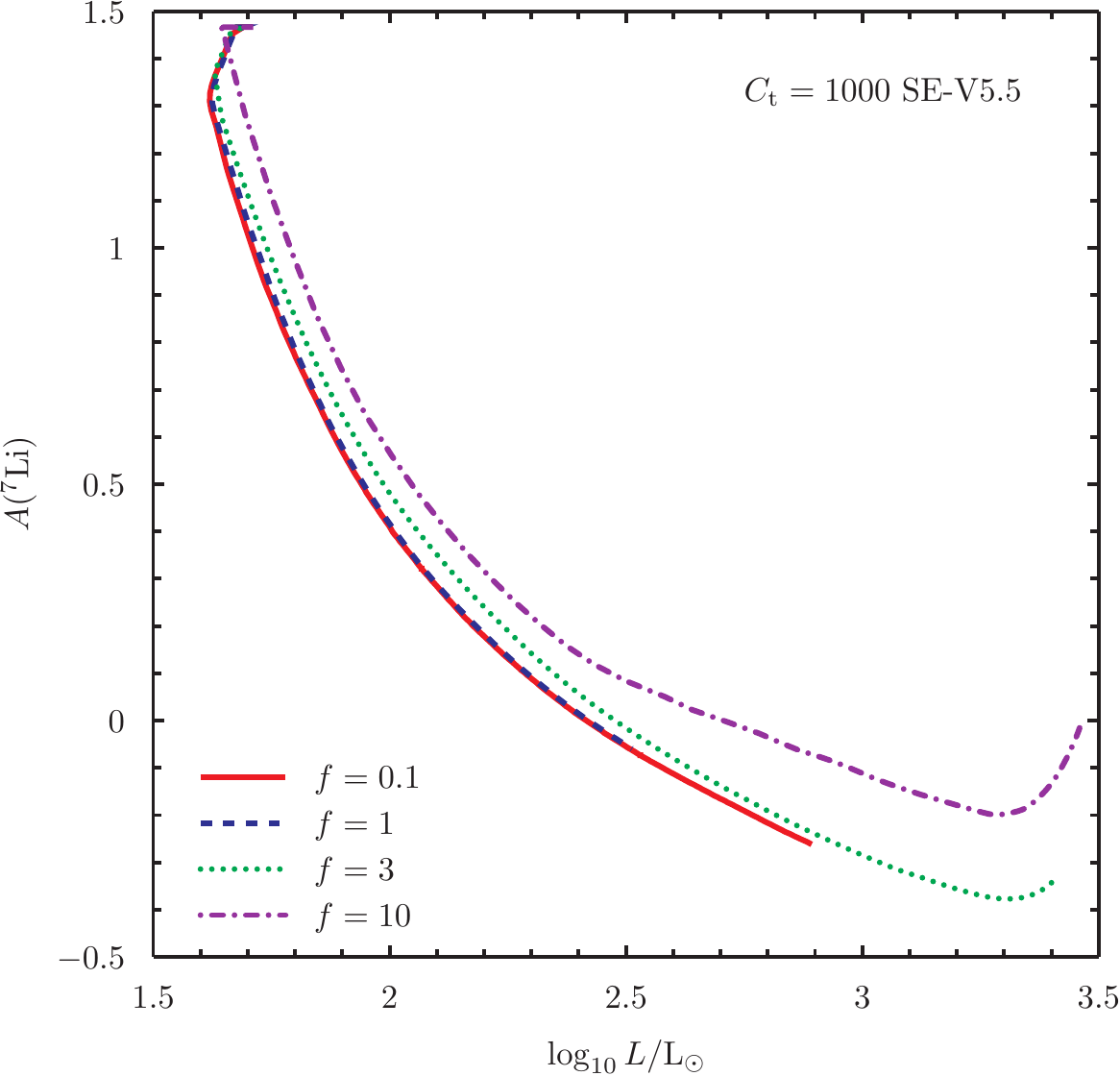}}
\caption{
Tests for the timestep criteria on the evolution of lithium abundance during the
giant branch.  $A({\rm Li})$ is plotted as a function of luminosity $L$.  The
mixing parameter $C_{\rm t}=1000$.  Different lines show different multiples $f$
of the preferred timestep $\delta t_0$ as defined in Eqn.~\ref{eqn:deltat0}.  The
solid red line shows $f=0.1$, the dashed blue line (under the solid red line)
$f=1$, the green dotted line $f=3$ and the purple dot-dashed line $f=10$.
}
\label{ftests}
\end{figure}

We return to our standard case with $C_t = 1000$ and run tests with {\sc se-v5.5} using a timestep
that is limited by $f \times \delta t_0$ with $f=0.1, 1.0, 3.0$ and $10.0$. Clearly $f$ is related to the
$a_i$ above, and we consider that it must be determined empirically. 
The results of these tests are shown in 
Figure~\ref{ftests}. We see that we appear to have converged on a solution for all except the case with $f=10$.
This confirms our claim that the important time is of order
$\tau_0$ and as long as the timestep stays near this value, 
the solution should be reliable. Note that for these tests we disabled  the other timestep
criteria; this means that for the $f=10$ case for example, we continued to use $\delta t = 10\tau_0$ even though
we would normally be using a smaller $\delta t$ as determined by another condition (eg $\delta t < 0.3\tau_{KH}$, as is 
normally used in {\sc se-v5.5}). Hence if a normal timestep condition produces a $\delta t$ that is below the
critical value, then naturally the solution is reached. But to be sure of this, we suggest adding a condition
that $\delta t$ not exceed say $3\tau_0$. This should ensure an accurate solution.

\begin{figure*}
\begin{center}
\includegraphics[width=0.4\textwidth]{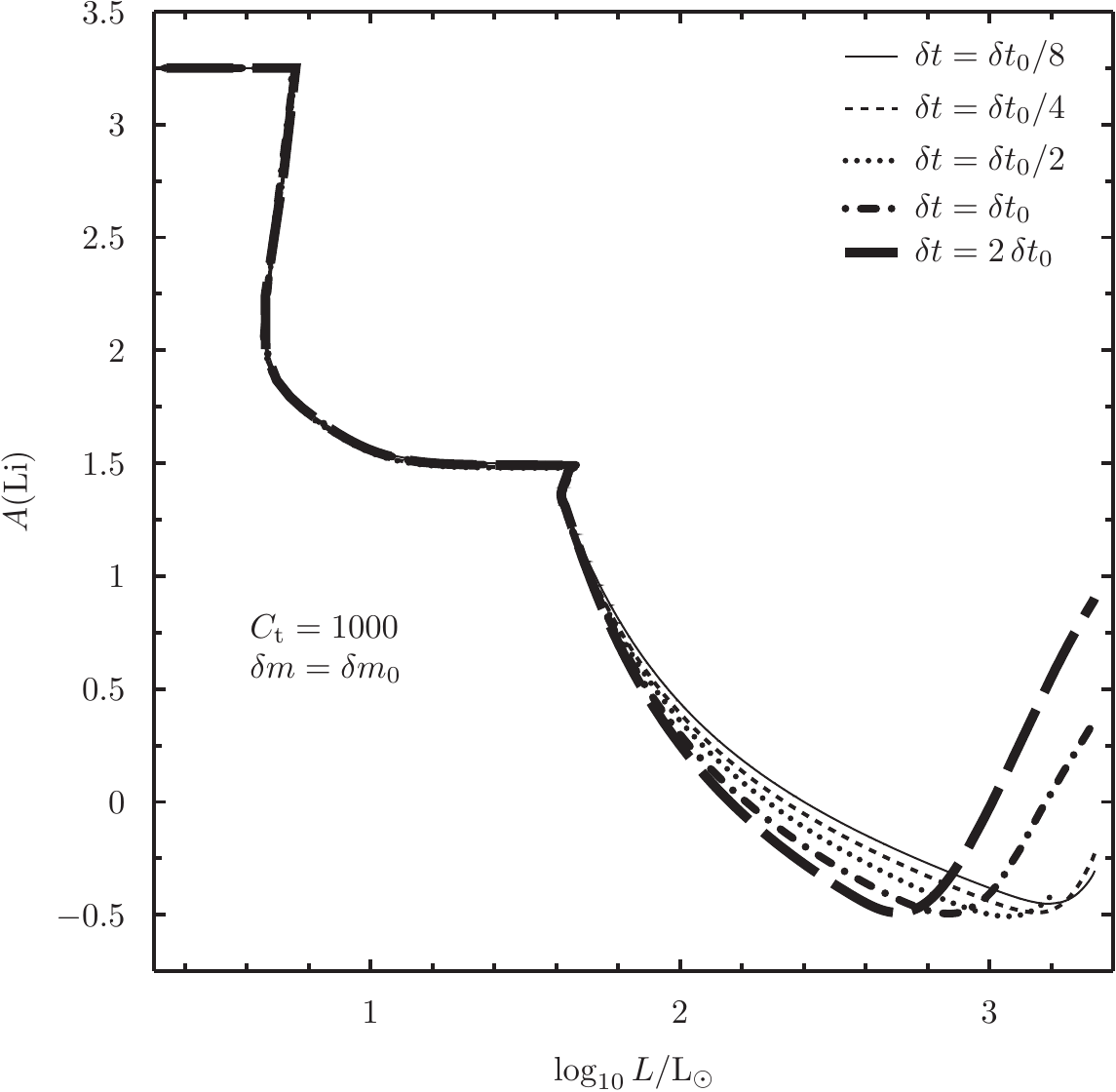}\hspace{2cm}
\includegraphics[width=0.4\textwidth]{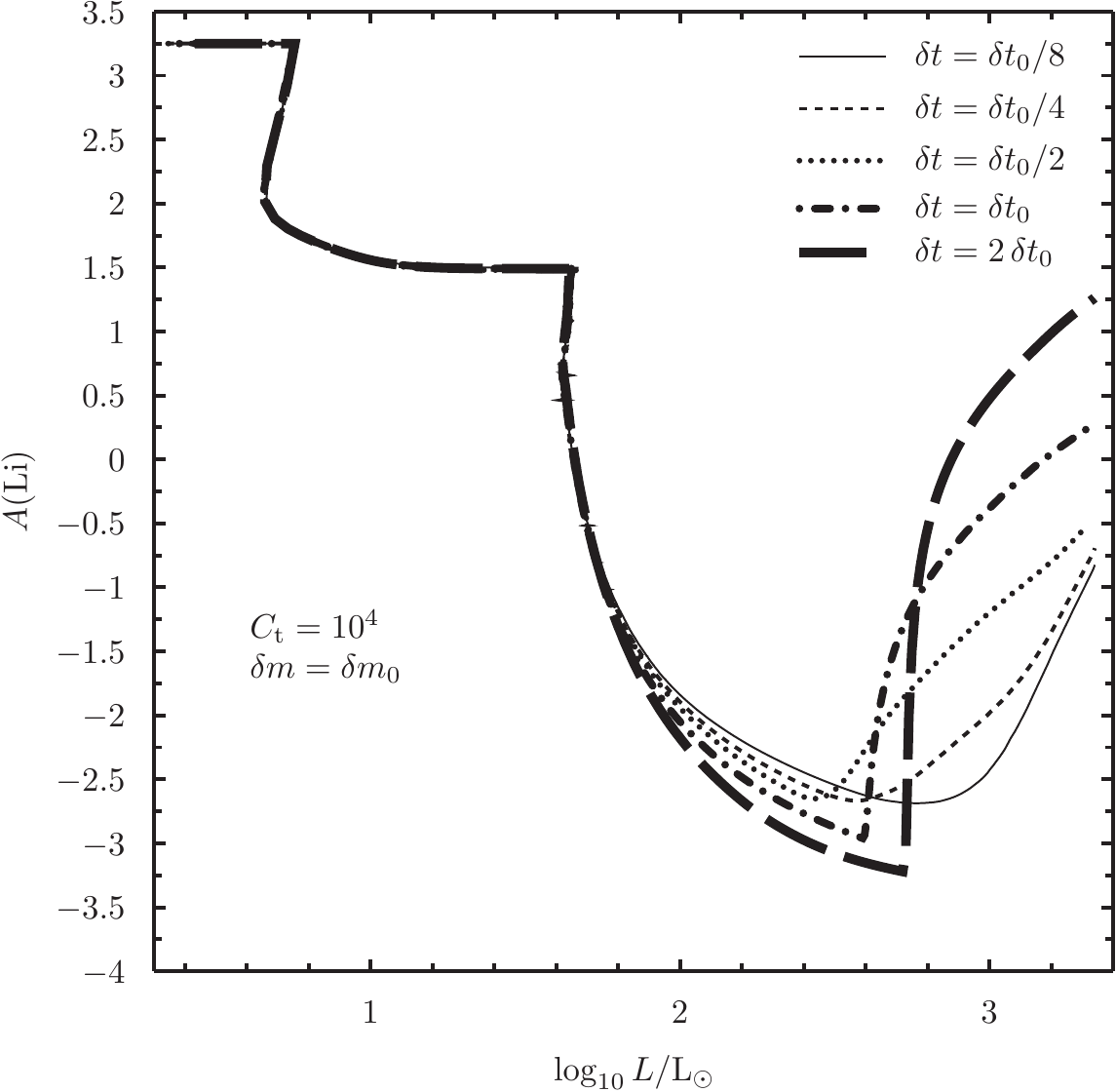}
\end{center}
\caption{
Tests of the effect of the timestep at fine mesh spacing, performed with the
{\sc monstar} code.  The lithium abundance $A({\rm Li})$ is plotted as a
function of luminosity $L$.  The left panel shows $C_{\rm t}=1000$, the right
panel $C_{\rm t}=10\,000$.  Each calculation used a maximum mesh spacing of $\delta
m_0$.  The lines show calculations with timesteps of $\delta t_0/8$ (solid thin
line), $\delta t_0/4$ (dashed thin line), $\delta t_0/2$ (dotted thin line),
$\delta t_0$ (dash-dotted thick line),  and $2\delta t_0$ (long-dashed thick
line).
}
\label{monstardttests}
\end{figure*}

We note that the {\sc se-v5.5} code solves simultaneously for the burning and mixing, whereas not all codes
do this; e.g. {\sc monstar} and {\sc se-g-v2.3} do not. 
Hence we perform tests also with {\sc monstar}\footnote{We have used existing
timestep controls in {\sc monstar} to simulate the implementation of $\delta t_0$, with the
result that our timestep is not precisely a multiple of $\delta t_0$, 
but is always between 0.5 and 2.0 times the quoted value.} 
to check if this affects
the preferred choice of $\delta t$. We anticipate slightly the next section, and perform these tests
with our preferred mesh spacing $\delta m_0$, although as we will show below, this is not crucial. 
Figure~\ref{monstardttests} shows 
results for $C_t=1000$ (left) and $10\,000$ (right) for various $\delta t$ as multiples of $\delta t_0$.
Although the early changes in Li content are not dependent on the timestep,  a converged
solution at later times requires a step no larger than about $\delta t_0/8$. This is about an order
of magnitude smaller than required by {\sc se-v5.5} and reflects the advantage of a simultaneous solution
for the burning and mixing. 

\begin{figure}
\begin{center}
\resizebox{0.48\textwidth}{!}{\includegraphics{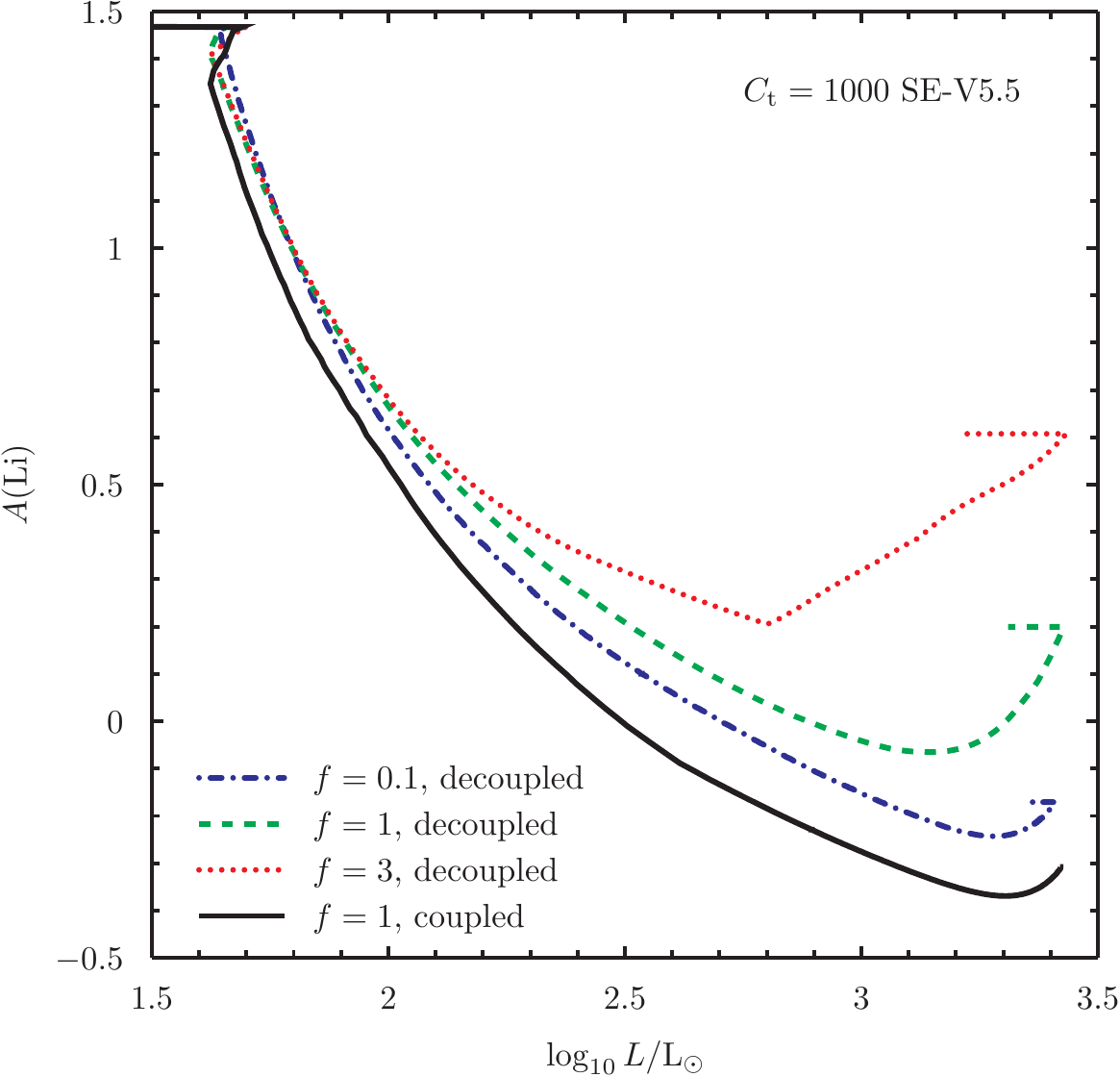}}
\end{center}
\caption{
Results from {\sc se-v5.5} with the calculations of mixing and burning of composition decoupled.
The blue dot-dashed line is for $f=0.1$, the green dashed line is for $f=1$ and
the red dotted line is for $f=3$.
The black curve shows the solution obtained with a simultaneous solution for 
mixing and burning and 
a timestep of $\delta t_0$ (i.e. $f=1$). All calculations use $C_{\rm t}=1000$.
}
\label{decoupled}
\end{figure}

To confirm this we repeated the calculations with {\sc se-v5.5}
with the mixing and burning algorithms decoupled, as is done in many codes,  
including {\sc monstar}
and {\sc se-g-v2.3}.
The results are shown in Figure~\ref{decoupled}, for tests with timestpes of $f \times \delta t_0$. 
Clearly to converge on the solution in this case requires
a value of $f \lesssim 0.1$, in agreement with the results from {\sc monstar}.
We conclude that our critical timestep is verified, but whether one should
use a limit of a few $\delta t_0$ or a few $\delta t_0/10$ depends on the solution scheme 
for the chemical composition that is used in the code in question.

Note also that it is imperative to apply this condition from the start of the thermohaline mixing; 
i.e. from the time when a $\mu$ minimum first emerges. 
If the timestep is too large initially then we follow a solution that is not appropriate. Using a 
smaller timestep later just ensures that we accurately follow an incorrect solution branch. 
There is no short cut -- a small $\delta t$ must be used from the start of the thermohaline mixing period.

\subsection{Mesh spacing}
\begin{figure*}
\begin{center}
\includegraphics[width=0.4\textwidth]{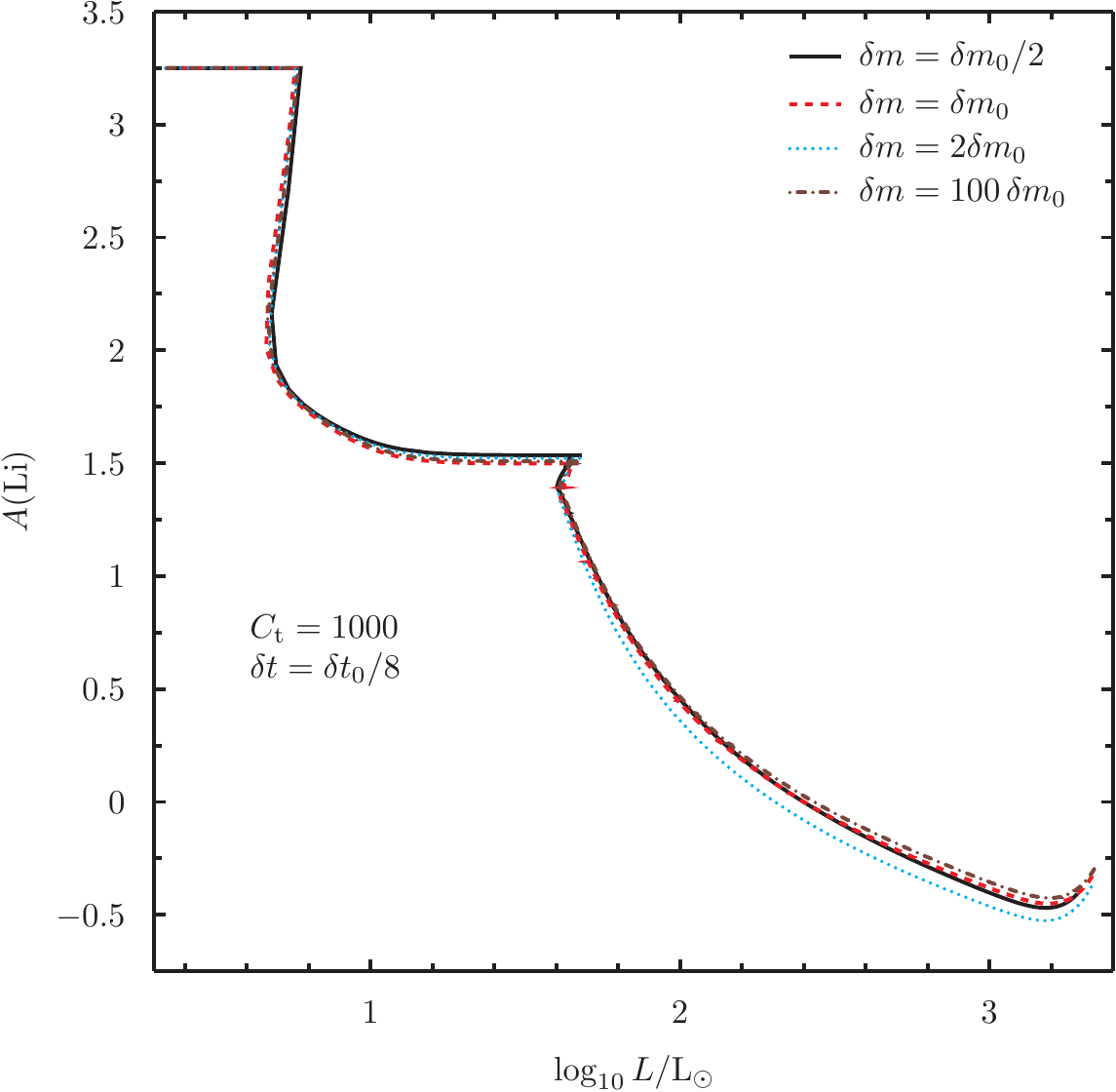}\hspace{2cm}
\includegraphics[width=0.4\textwidth]{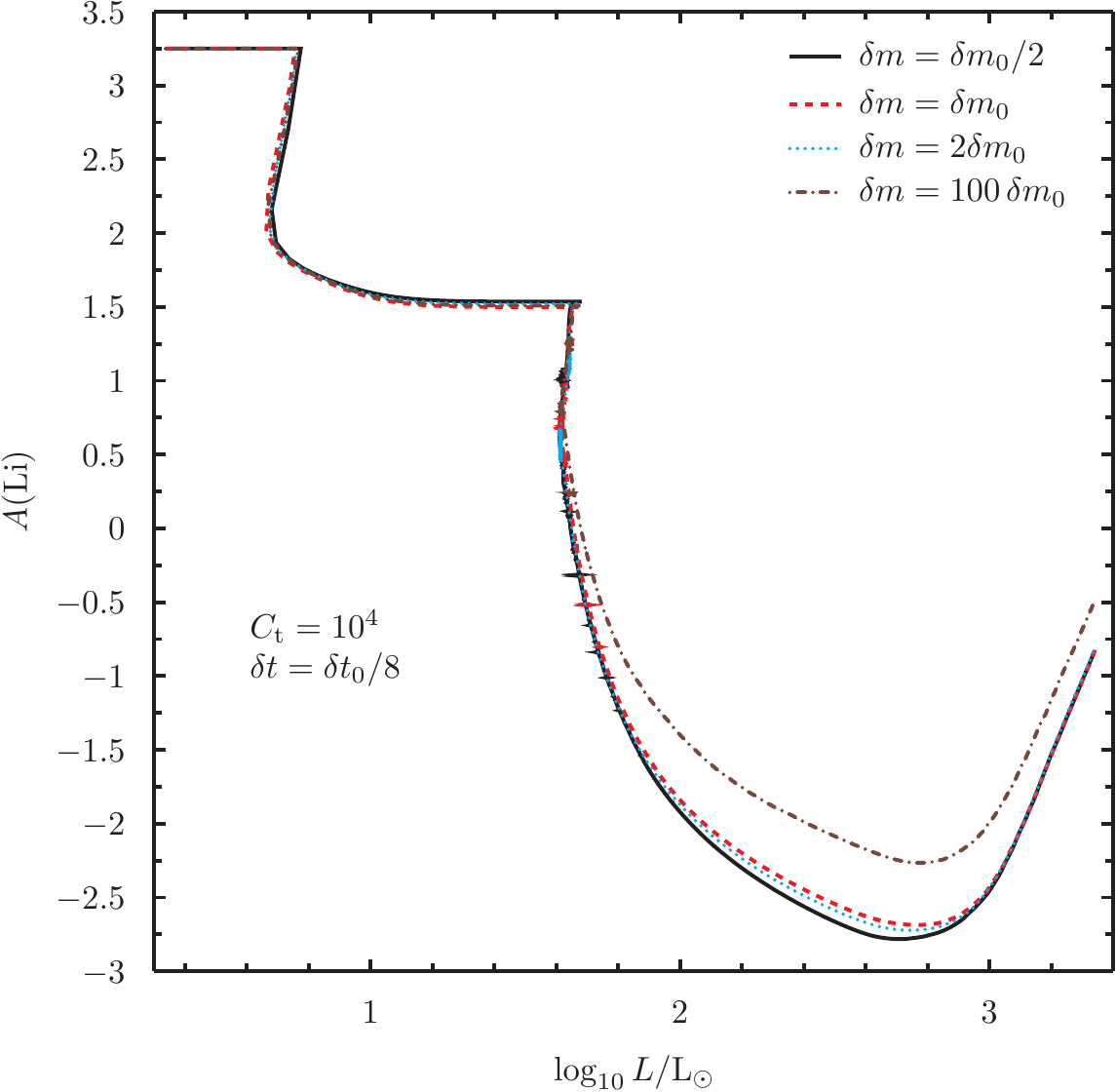}
\end{center}
\caption{
Tests of the effect of the mesh spacing performed with the {\sc monstar} code.  The
lithium abundance $A({\rm Li})$ is plotted as a function of luminosity $L$.  The
left panel shows $C_{\rm t}=1000$, the right panel $C_{\rm t}=10\,000$.
Each calculation used a maximum timestep of $\delta t_0/8$.  The lines show
calculations with mesh spacing of $\delta m_0/2$ (solid black line), $\delta m_0$
(dashed red line), $2\delta m_0$ (dotted blue line)
and $100\delta m_0$ (dash-dotted brown line).
}
\label{monstardmtests}
\end{figure*}

We also performed tests on the mesh spacing used in the calculations. 
Because the thermohaline mixing is being driven by variations in $\mu$, which are
in turn driven by the burning of $^3$He, we concentrate on fully resolving variations in the
$^3$He content. Based on experience with
such problems, we defined a preferred maximum mass spacing $\delta m_0$ so that the maximum 
change in the mass fraction of $^3$He between mesh points is no more than 1\% of the maximum 
value of the $^3$He  abundance throughout the star. We performed the tests with {\sc monstar} and the 
standard model, with $C_t = 1000$ and the very sensitive case of  $C_t = 10\,000$.
The results are shown in 
Fig~\ref{monstardmtests}, where we have used our preferred timestep of $\delta t_0/8.$ 
These plots clearly show that if the timestep is short enough, 
then the mesh spacing is not critical, although it is more important in the $C_t=10\,000$ case, 
as may be expected. But even here, we have to use $100\delta m_0$ to produce a significant degradation of the results.

\subsection{Combining Space and Time Resolution}\label{sec-combined}
The choices of space and time resolution are in fact related. For a sufficiently poor resolution in one variable it
is unlikely that the correct solution will be recovered for any choice of the other  variable. Here we test
the interdependence of the spatial and temporal resolution criteria. Again, tests were performed 
with the {\sc monstar} code and for the two cases $C_t=1000$ and  $C_t=10\,000$. For each we have
varied the timestep significantly for two choices of the mesh spacing: $\delta m_0$ and $100\delta m_0$.

Let us first examine the case for $C_t=1000$ and $\delta m_0$, as shown in the left panel of Figure~\ref{monstardttests}.
Clearly the solution converges as the timestep is lowered, and with a maximum $\delta t = \delta t_0/8$ we 
believe we have essentially found the solution. With larger timesteps we obtain
values that
are quite different to the true solution, although the variations are not large until
we begin producing Li toward the tip of the RGB (see \S\ref{secLirise} for a discussion of this phenomenon).
For the case with $C_t = 10\,000$ in the right panel of Figure~\ref{monstardttests} the runs
with larger $\delta t$ are clearly wrong, with large discontinuities produced in the
surface Li abundance. As the timestep is decreased however we again appear to converge on
a solution.

\begin{figure*}
\begin{center}
\includegraphics[width=0.4\textwidth]{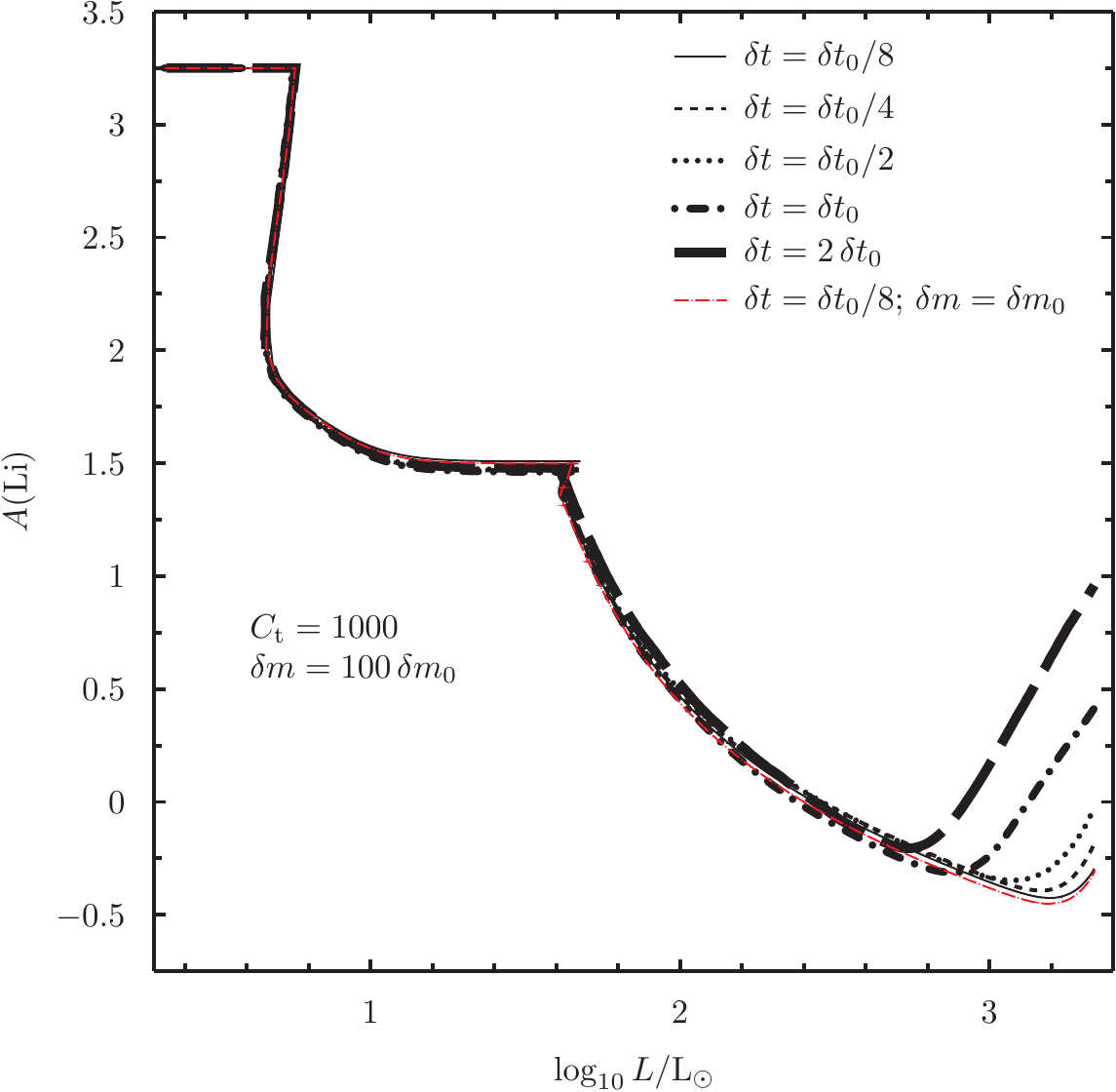}\hspace{2cm}
\includegraphics[width=0.4\textwidth]{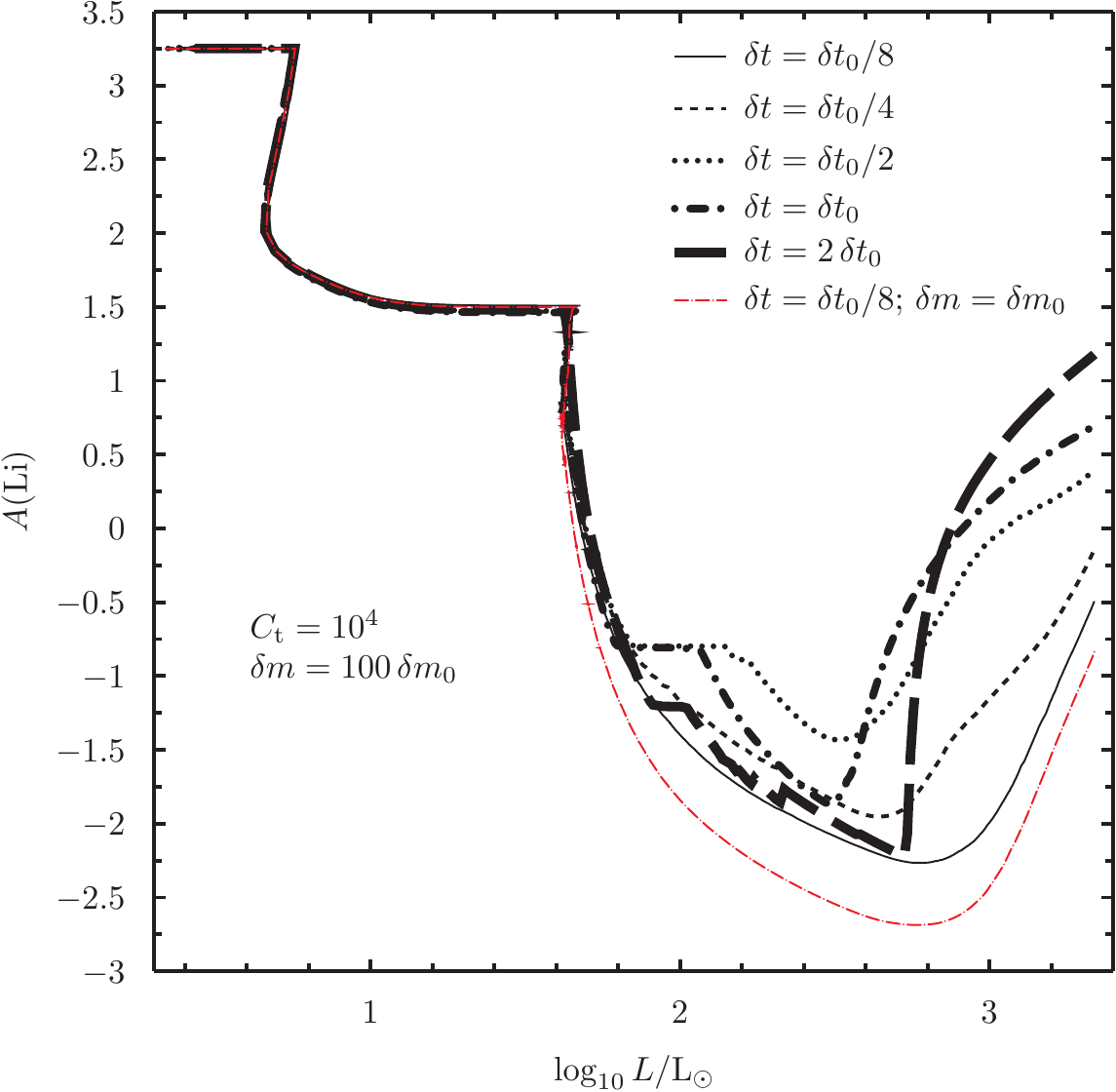}
\end{center}
\caption{
Tests of the effect of the timestep at coarse mesh spacing, performed with the
{\sc monstar} code.  The lithium abundance $A({\rm Li})$ is plotted as a
function of luminosity $L$.  The left panel shows $C_{\rm t}=1000$, the right
panel $C_{\rm t}=10\,000$.  Each calculation used a maximum mesh spacing of
$100\,\delta m_0$.  The lines show calculations with timesteps of $\delta t_0/8$
(solid thin line), $\delta t_0/4$ (dashed thin line), $\delta t_0/2$ (dotted
thin line), $\delta t_0$ (dash-dotted thick line),  and $2\delta t_0$ (long-dashed thick line).
}
\label{monstardttests100dm}
\end{figure*}

We now wish to determine how sensitive is this result to the mesh spacing. Hence we repeat the
two tests above, but using a maximum mesh spacing of $\delta m = 100\delta m_0$, as shown in Figure~\ref{monstardttests100dm}.
Firstly we look at the left panel, for $C_t = 1000$. Again, as we decrease the timesteps the
solution converges. What is interesting however is the red line, which is the converged solution using
a mesh spacing that is 100 times smaller. This solution 
with $\delta t = \delta_0/8$
matches the converged solution, even for the
case with $\delta m = 100\delta m_0$. i.e. the thin red line and the thin black line are essentially the same.
This tells us that it is the timestep that is by far the most important parameter in finding the solution to this
problem. Provided the timestep is small enough, the mesh spacing is not critical, to within relatively large factors.
The same thing is seen in the right panel, for $C_t = 10\,000$. Here, with the larger timesteps there are
discontinuities and large changes in the solution.
Decreasing the 
timesteps removes the erratic behaviour and produces a smooth solution. Here again the red curve is the solution
using 100 times smaller mesh spacing. In this more sensitive case there is still a difference between the
solutions for the smallest timestep, so we conclude that in the extreme cases the mesh spacing is indeed important,
although the crucial thing to be concerned with is the timestep chosen. The mesh spacing is important but
not as critical, except in the most strongly mixed cases.

\begin{figure*}
\begin{center}
\includegraphics[width=0.4\textwidth]{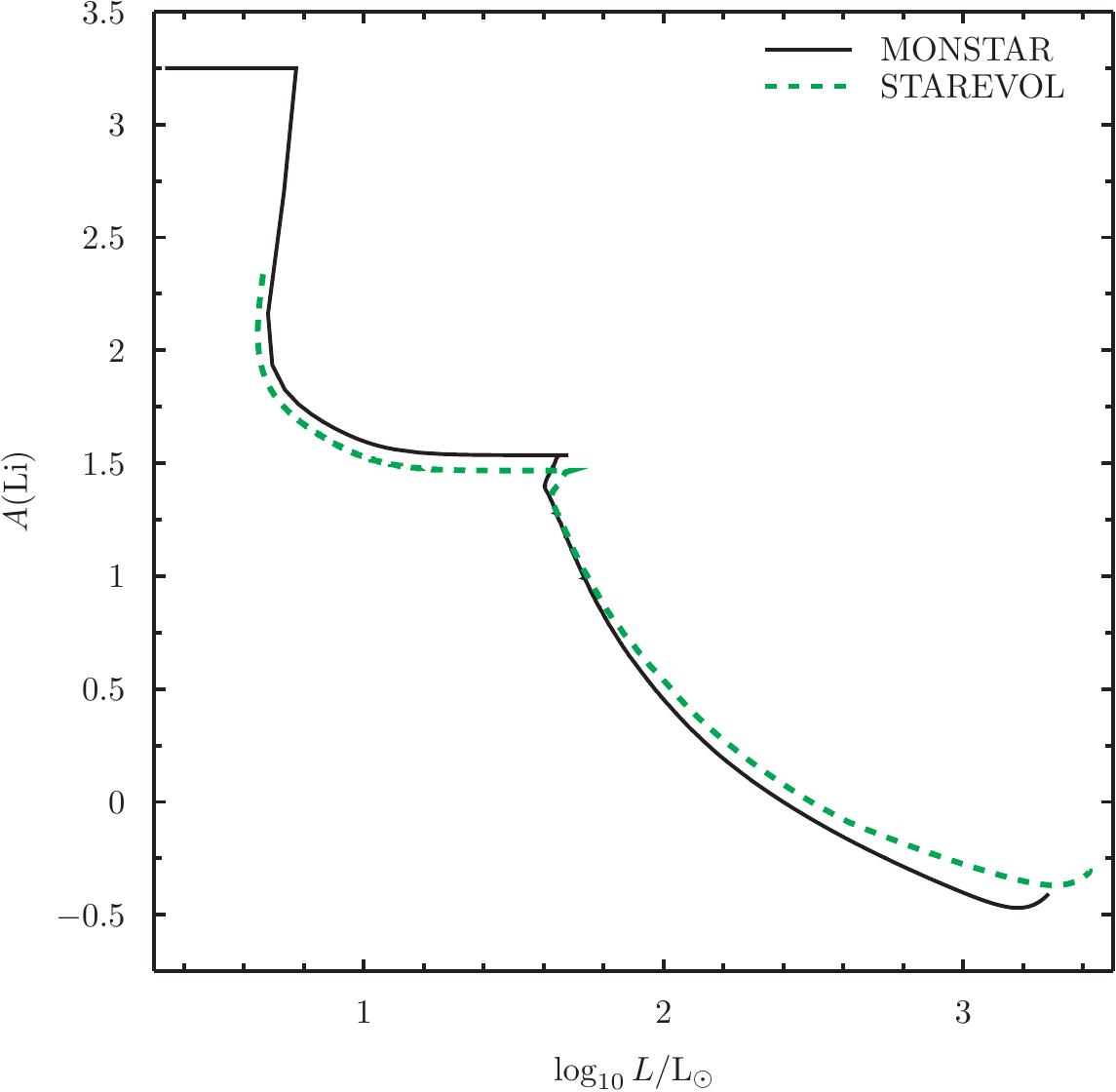}\hspace{2cm}
\includegraphics[width=0.4\textwidth]{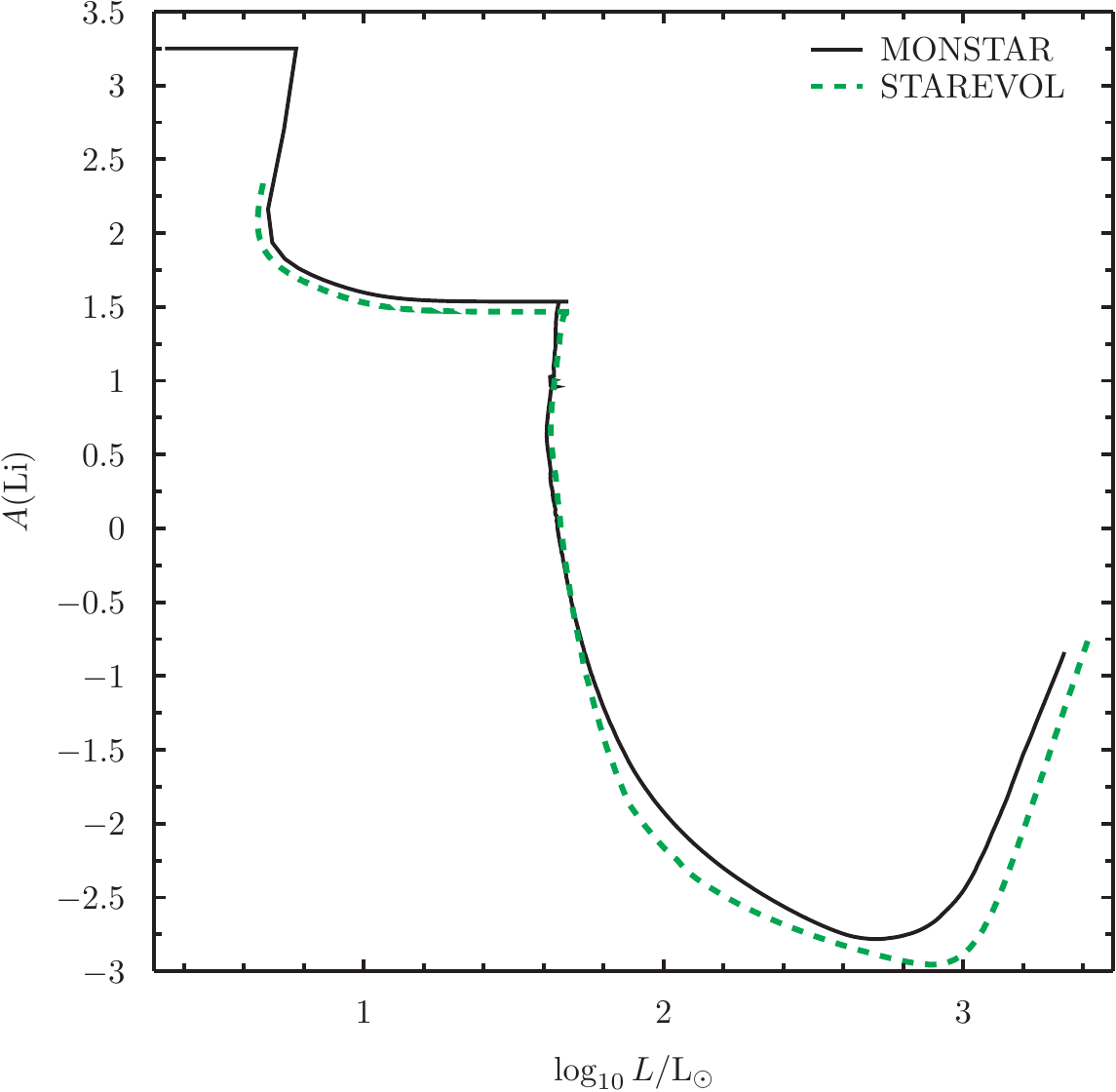}
\end{center}
\caption{Calculations with our preferred time and mesh spacing. 
The lithium abundance $A({\rm Li})$ is plotted as a function of luminosity $L$.
The left panel shows $C_{\rm t}=1000$, the right panel $C_{\rm t}=10\,000$.  Solid
black lines are results from {\sc monstar}; dashed green lines results from {\sc
se-v5.5}.
}
\label{converged}
\end{figure*}

Finally, to check that our preferred mesh and time spacing works, we show in Figure~\ref{converged} 
the results for $C_t=1000$ and $C_t=10\,000$ using both {\sc monstar} and {\sc se-v5.5}. 
Both codes use $\delta m = \delta m_0$, while {\sc se-v5.5} uses $\delta t = \delta t_0$ and
{\sc monstar} uses $\delta t = \delta t_0/8$, as determined earlier.
We see
that they do in fact agree sufficiently, and thus verify that the critical solution mesh proposed in this
paper is suitable for calculations of the behaviour of Li during the RGB phase, in different evolution codes.

We have not tested our proposed solution mesh in the {\sc stars} code. This code uses an elegant implicit 
method to solve for its preferred solution mesh. This does not lend itself to the sort of tests
and controls we derived here, so we do not try to apply them to this code. Nor have we implemented our
preferred mesh in {\sc mesa}. We have instead shown that our suggested solution mesh works well in two
totally independent codes. We advise anyone who is interested in calculating thermohaline mixing
of Li to perform tests of the code being used, guided by our preferred $\delta t_0$ and $\delta m_0$.

\section{The Production of Lithium}\label{secLirise}
We have seen that the models show a tendency to produce Li as they get near to the tip of the giant branch.
This tendency is stronger as $C_t$ increases, being essentially absent when $C_t=100$ but almost universal 
when $C_t=10\,000$ (the only exception being for the {\sc se-g-v2.3} code). Our aim in this
section is to understand why this happens.
That deep-mixing can produce Li is not a new result, having been discussed extensively by \citet{SB96Li}. The question
is how and why do our models change from efficient destruction of Li early on the RGB to efficient production
later on the RGB. We defer to a companion paper (Church et al 2014, in preparation) the question about whether real 
stars behave this way; our emphasis here is
on the accurate solution of the numerical problem rather than a test of its applicability.

\begin{figure}
\resizebox{0.49\textwidth}{!}{\includegraphics{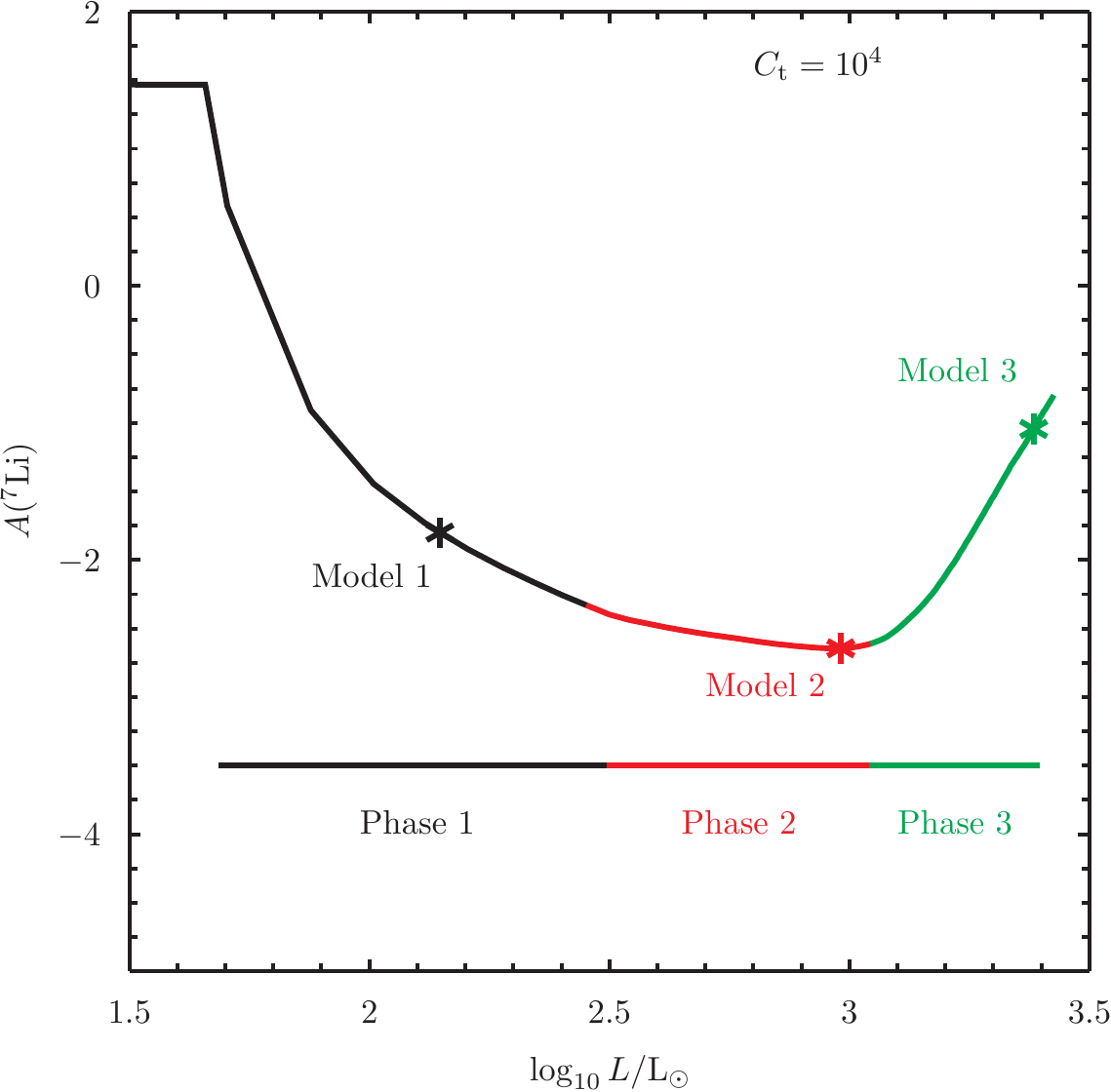}}
\caption{
The three phases of different Li behaviour highlighted in different colours. The
asterisks indicate the specific models plotted in Figures~\ref{Lirise} and \ref{Drise}.
}
\label{phases}
\end{figure}

Here we look at the case with $C_t = 10\,000$ since this is where the Li production appears most easily. We show the
results from {\sc se-v5.5} in Figure~\ref{phases} and identify three phases of the evolution. In Phase~1 we 
see a decrease in the envelope Li content, which reaches 
a minimum at Phase~2, and then during Phase~3 the Li increases again.

\begin{figure*}
\resizebox{0.7\textwidth}{!}{\includegraphics{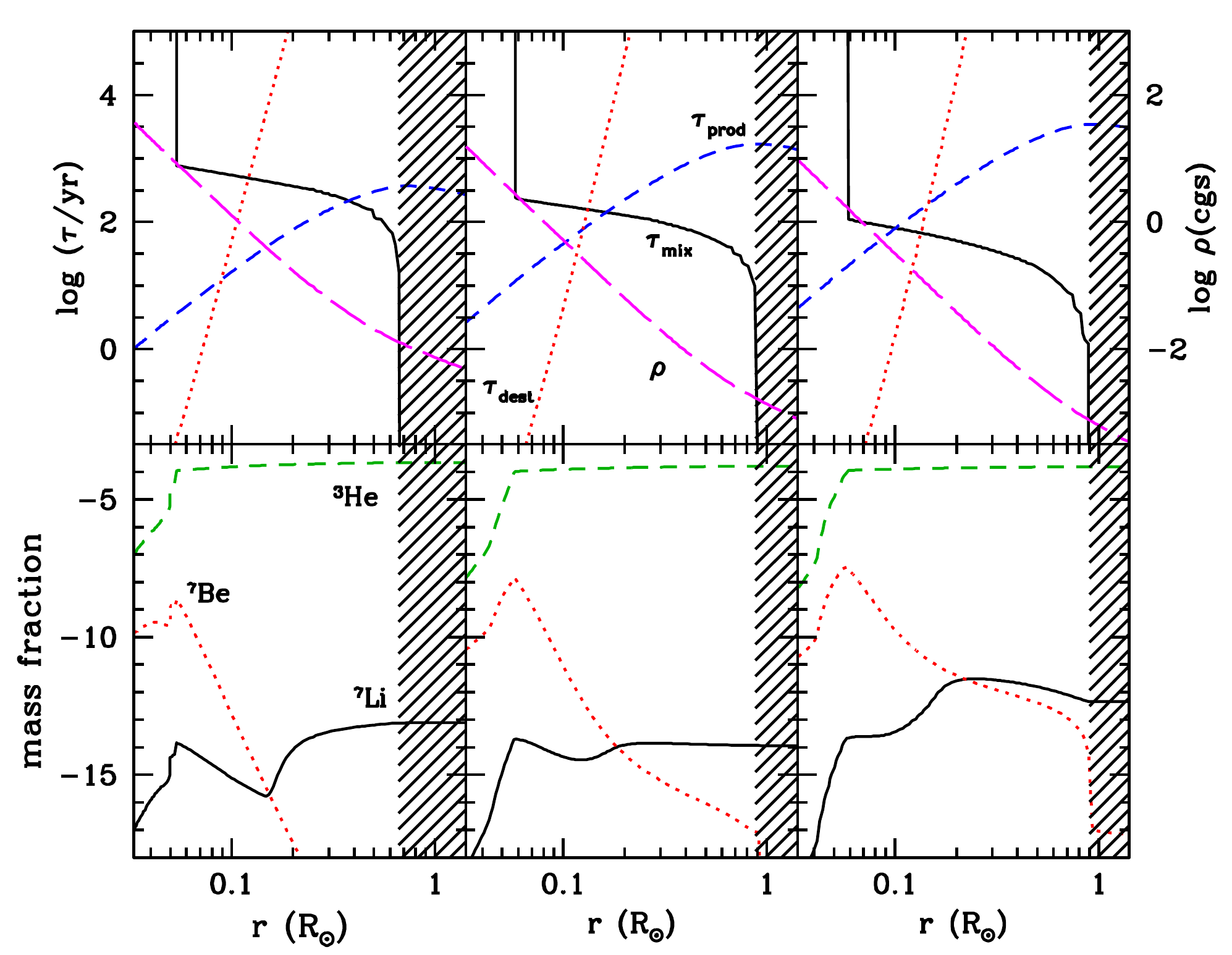}}
\caption{Structure in the three models specified in Figure~\ref{phases}. Model 1
(left) is during the phase of evolution where thermohaline mixing reduces the
surface lithium abundance. Model 2 (middle) is where the surface lithium
abundance is roughly constant.  Model 3 (right) is from near the tip of the
giant branch where the surface lithium abundance is increasing.
The top panels show the important timescales and the density.  
The dashed blue lines show the timescale
for the production of $^7{\rm Li}$, the dotted red lines show the timescale for its
destruction.  The solid black line shows the timescale to mix from a given point to
the base of the convective envelope.  The magenta line shows the run
of density $\rho$ in units of g\,cm$^{-3}$.
The bottom panels show composition for selected species  The solid black line is
$^7{\rm Li}$, the red dotted line is for $^7{\rm Be}$ and the green dashed line 
is for $^3{\rm He}$.
}
\label{Lirise}
\end{figure*}

The left, middle and right panels in Figure~\ref{Lirise} correspond to the 
three models denoted by asterisks in Figure~\ref{phases}. The upper panels show three 
timescales: 
\begin{itemize}
\item $\tau_{prod}$ (blue line) which is the timescale for the production of \chem{7}Li, which is also 
the timescale for the destruction of \chem{7}Be (since the rate of proton capture on \chem{7}Be is negligible); 
\item $\tau_{dest}$ (red line) which is the timescale for the destruction of \chem{7}Li;
\item $\tau_{mix}$ (black line) which is the timescale for the matter to diffuse from the 
given position to the bottom of the 
convective envelope;
\end{itemize}

\noindent and the lower panels show the abundance profiles of $^7$Li, $^7$Be and $^3$He.

In the bottom left panel we show the situation typical of phase~1. The Li content in the convective envelope 
(shown hashed) is higher than in the 
interior and Li will diffuse inward. Just below the envelope we see a region where the production of Li is
much faster than the destruction timescale. However this production is negligible compared to the 
amount of Li transported from the envelope reservoir.
The production and mixing contribute mostly to filling in the little dip in Li seen at $r \simeq 0.15$\rsun. 
A little further 
interior we see that Li destruction is the dominant process, and indeed it is the transport of Li 
from the surface into this region that leads to a reduction in its surface abundance.
Note that there is a region 
of enhanced \chem{7}Be which is produced by $\alpha$ captures on \chem{3}He.
The destruction of \chem{7}Be is only efficient at the very bottom of the thermohaline
region ($r \la 0.06 R_\odot$) where we see a decrease in the \chem{3}He \chem{7}Be abundance profiles.
As the other panels show, overall
the \chem{7}Be content increases during the rise up the RGB.

In the middle panel the Li profile has flattened, and hence the diffusion inward must cease. We have reached a 
local extremum (minimum) in the surface Li abundance. From the upper middle panel we see that the nuclear timescales
have hardly changed, whereas the mixing timescale has shortened compared to phase~1. We also see the
\chem{7}Be start to diffuse outward from where it was produced.

This decrease in the mixing timescale continues into phase~3, as shown in the right top panel. Now the increase
in \chem{7}Be below the envelope results in a production of \chem{7}Li in this region
due to electron capture reactions. \chem{7}Li is produced 
in the region at $r \simeq 0.1$--$0.5$\rsun, and it is efficiently transported into the envelope causing the surface 
\chem{7}Li content to rise.

In summary, the main change in the structure as the star ascends the RGB is that the diffusion becomes
much quicker, allowing the \chem{7}Be to escape from where it is produced. This is the classical 
Cameron-Fowler mechanism in operation.

\begin{figure}
\resizebox{0.4\textwidth}{!}{\includegraphics{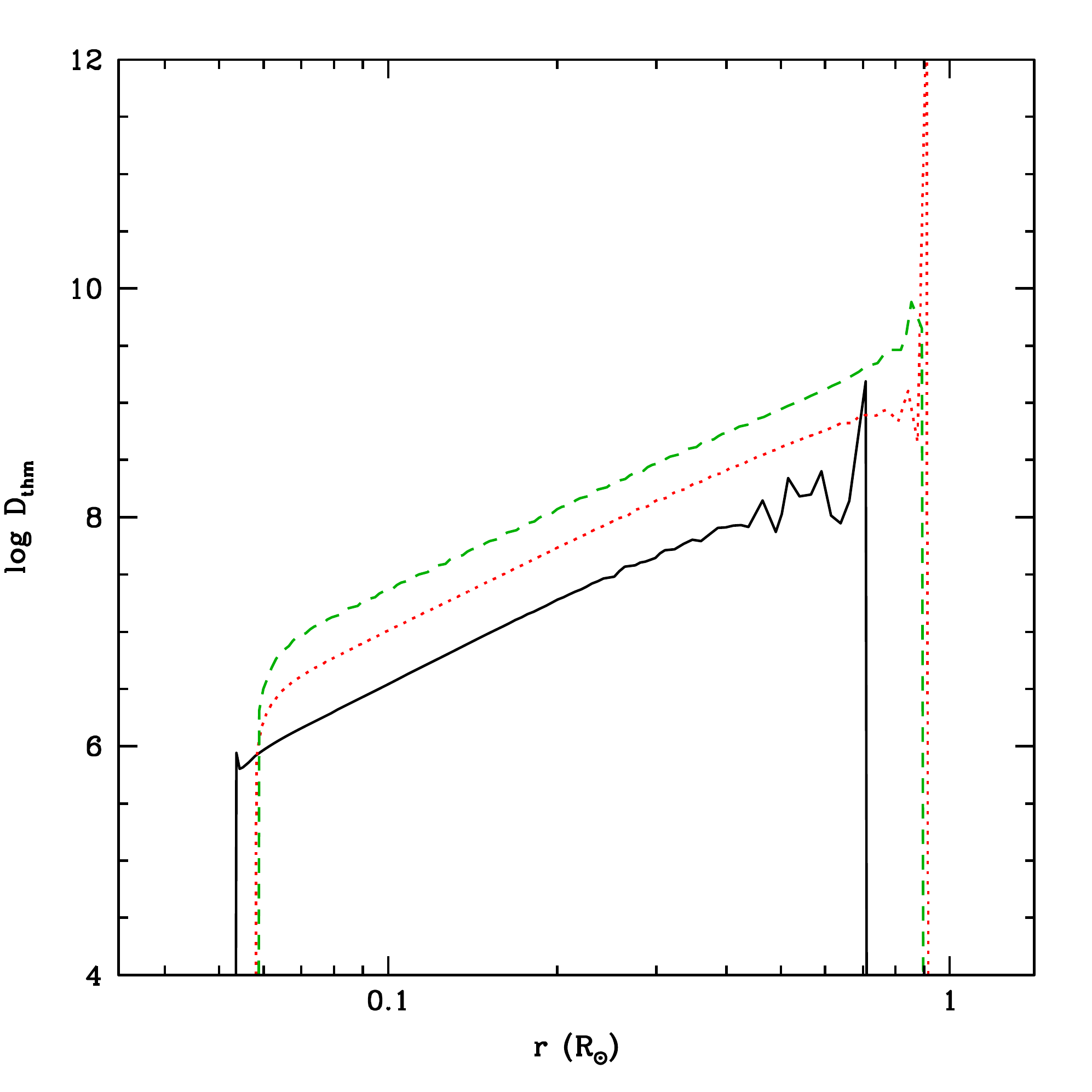}}
\caption{
Variation of the thermohaline diffusion coefficient $D_{\rm thm}$ as a function
of radius $r$ for the three models shown in Figures~\ref{phases}
and~\ref{Lirise}: Model 1 (surface lithium reducing) is shown by the  solid
black line, Model 2 (surface lithium constant) by the dotted red line, and Model
3 (surface lithium increasing) by the dashed green line.  The roughness in the
curve near the base of the convective envelope is caused by the small value of
$\nabla\mu$ in this region.}
\label{Drise}
\end{figure}

It remains for us to understand why the mixing timescale decreases as the 
model ascends the RGB. The mixing timescale is just the time taken to diffuse from
a given radius to the bottom of the envelope. This is the sum of the terms $(\delta r)^2/D_{thm}$ for 
each shell across the region, emphasizing again the reason why we have plotted our graphs against 
radius instead of mass. The spatial extent of this region stays essentially the same, as shown in 
Figure~\ref{Lirise}. (Importantly, the mass in this region 
decreases from $9.38\times 10^{-3}$\msun through $3.45\times10^{-3}$\msun to $1.93\times 10^{-3}$\msun for 
the three models shown in Figure~\ref{Lirise}.) Hence we expect that $D_{thm}$ increases as the star ascends the RGB, 
and this is indeed what we find, as shown in Figure~\ref{Drise}. 

The expression for $D_{thm}$ is given in equation~(1). We have investigated each term in this
expression for $D_{thm}$ to determine which is responsible for the increase. The dominant
term is easily the thermal diffusivity $K$, and within $K$ the largest change is due to the
decrease in density as the star expands along the RGB. This can be understood as altering the
fundamental physical structure of the region. As the thermal diffusivity increases then the heat can be transported
very easily, making the material with a lower $\mu$ even more buoyant, and increasing the
diffusive motions. 

In fact any changes that result in an expansion in this region would act to favour 
Li production. Since the strength of the hydrogen burning shell scales directly with the initial 
CNO abundance, we would expect that Li production would be more favoured
in higher metallicity stars.

\section{Summary and Conclusions}

Lithium continues to cause problems for many branches of astrophysics. At the very least we need
to understand how it behaves in stars. We have seen in this paper that even within the constraints of a well
defined mathematical model for diffusive mixing we find dramatically different  behaviours
due to the high sensitivity of Li to physical conditions in the star. We find that 
although the dominant effect of thermohaline mixing is to decrease the
stellar surface content of Li on the RGB, the Cameron-Fowler mechanism
can operate near the tip of the RGB for sufficiently high values of the parameter $C_t$ that appears in the 1D
diffusion theory for thermohaline mixing. Further, unless sufficient care is taken in the integration
of the diffusion equation and the structure equations, one can easily find 
envelope lithium abundances that differ from a resolved solution
by orders of magnitude. We present criteria to be used for
determining the timesteps and spatial resolution needed for an accurate solution of the Li content during thermohaline mixing.
These criteria can be summarised as:
\begin{enumerate}
\item the timestep $\delta t$ should satisfy
\begin{equation}
\delta t \lesssim f \times \sum_i \frac{(\delta r_i)^2}{D_i}
\end{equation}
where the sum is taken from the point where the Li production and 
destruction timescales are equal, to the bottom of the convective envelope;
$f\simeq 2$ is suitable for an implicit and simultaneous solution of burning and 
mixing, but if these processes are calculated separately then $f\lesssim 0.2$ is required;
\item the spatial mesh spacing $\delta m$ should be no larger than will permit 
a change in the $^3$He abundance of a few percent of the maximum $^3$He value in the model.
\end{enumerate}
Of these, the timestep criterion is the most important, in that if the behaviour is resolved in time then the
spatial mesh is not crucial, except for  higher values of $C_t$.
We show that by using these criteria we find the results produced by two totally independent codes are in agreement.
It would be prudent for anyone interested in this problem to perform
similar tests with the code they will be using, guided by our recommended mesh spacings.

Finally we note that any mechanism that will determine the composition of fragile elements like Li is likely to be 
very sensitive to numerical details of the kind discussed here. Hence an analysis similar 
to that performed here would be wise when investigating any other proposed mechanism, 
especially one that has feedback between the composition profile and the details of the mixing.

\section{Acknowledgments}
This research was supported under Australian Research Council's Discovery Projects 
funding scheme (project numbers DP0877317, DP1095368 and DP120101815). 
RPC is supported by the Swedish Research Council (grants 2012-2254 and
2012-5807). RJS is the recipient of a Sofja Kovalevskaja Award from the Alexander von Humboldt Foundation.
L.S. is research associate at the F.R.S.-FNRS.

{}

\end{document}